%% file: manuscript.tex
\documentclass[apsprl,twocolumn,superscriptaddress]{revtex4-2}

\input{source/preamble}

\defaultbibliographystyle{apsrev4-2}

\begin{document}

\input{source/main}
\input{source/supplements}
\end{document}

%% file: source/preamble.tex
\usepackage{graphicx,subfigure}
\usepackage{changes}
\usepackage{amsmath,amssymb}
\usepackage{mathtools}
\usepackage{multirow}
\usepackage{textcomp}
\usepackage{float}
\usepackage{xcolor}
\usepackage{titletoc}
\usepackage{ulem}
\usepackage{dsfont}
\usepackage{siunitx}
\DeclareSIUnit\gauss{G}
\usepackage{bibunits}
\usepackage{import}
\usepackage{braket}
\usepackage{nicefrac}
\usepackage{comment}
\usepackage{bm}
\usepackage{bbm}
\usepackage[english]{babel}
\usepackage{hyperref}

\usepackage[thinc]{esdiff}

\hypersetup{colorlinks=true,linkcolor=blue,citecolor=blue,breaklinks}

\newcommand{\expect}[1]{\ensuremath{\left\langle{#1}\right\rangle}}

\newcommand{\abs}[1]{\ensuremath{\left\vert{#1}\right\vert}}

\newcommand{\micro}[1]{\ensuremath{\mu\mathrm{#1}}}

\renewcommand{\micro}[1]{\ensuremath \mu\mathrm{#1}}

\newcommand{\dimH}{\ensuremath{\left(\mathrm{dim}\,\mathcal{H}_{\mathrm{eff}}\right)^{1/L}}}

\let\oldsfdefault\sfdefault
\usepackage[scaled=0.92]{helvet}
\renewcommand{\sfdefault}{\oldsfdefault}

\usepackage{layouts}

%% file: source/main.tex
\begin{bibunit}
\input{source/title}

\date{\today}


\begin{abstract}
The Loschmidt echo - the probability of a quantum many-body system to return to its initial state following a dynamical evolution - generally contains key information about a quantum system, relevant across various scientific fields including quantum chaos, quantum many-body physics, or high-energy physics. However, it is typically exponentially small in system size, posing an outstanding challenge for experiments. Here, we experimentally investigate the subsystem Loschmidt echo, a quasi-local observable that captures key features of the Loschmidt echo while being readily accessible experimentally. Utilizing quantum gas microscopy, we study its short- and long-time dynamics. In the short-time regime, we observe a dynamical quantum phase transition arising from genuine higher-order correlations. In the long-time regime, the subsystem Loschmidt echo allows us to quantitatively determine the effective dimension and structure of the accessible Hilbert space in the thermodynamic limit. Performing these measurements in the ergodic regime and in the presence of emergent kinetic constraints, we provide direct experimental evidence for ergodicity breaking due to fragmentation of the Hilbert space. Our results establish the subsystem Loschmidt echo as a novel and powerful tool that allows paradigmatic studies of both non-equilibrium dynamics and equilibrium thermodynamics of quantum many-body systems, applicable to a broad range of quantum simulation and computing platforms.
\end{abstract}
\maketitle

Analogue and digital quantum simulators, based on neutral atoms~\cite{gross_quantum_2021}, Rydberg atoms~\cite{browaeys_many-body_2020}, superconducting qubits~\cite{kjaergaard_superconducting_2020}, or trapped ions~\cite{foss-feig_progress_2024}, are powerful devices for exploring the inherently complex non-equilibrium dynamics of quantum many-body systems. To fully harness their potential, it is essential to identify practical resource-efficient observables that go beyond simple measurements of, e.g., local mean densities or low-order correlations. One particularly valuable quantity, which generally contains important information about the quantum many-body state and its dynamics, is the Loschmidt echo (LE). Also known as the return probability or fidelity, the LE quantifies the overlap between initial and time-evolved many-body states~\cite{footnote_Loschmidt_echo}
\begin{equation}
    \begin{aligned}
        \mathcal{L}(t) = |\braket{\psi_0|\psi(t)}|^2.
    \end{aligned}
    \label{eq:fullLE}
\end{equation}

The concept of return probability has a rich tradition in theoretical studies going back to Boltzmann and Poincar\'e~\cite{coveney_second_1988}, the theory of quantum chaos~\cite{peres_stability_1984, gorin_dynamics_2006, haake_quantum_2018}, random matrix theory~\cite{torres-herrera_dynamical_2017-1}, lattice gauge theories~\cite{schwinger_gauge_1951, cohen_schwinger_2008}, or large-deviation theory~\cite{touchette_large_2009}. More recently, the LE became instrumental in theoretical investigations of strongly-correlated many-body systems, such as many-body localization~\cite{torres-herrera_dynamics_2015, serbyn_loschmidt_2017}, many-body scarring~\cite{turner_weak_2018, serbyn_quantum_2021}, or even quantum metrology~\cite{li_improving_2023, colombo_time-reversal-based_2022, davis_approaching_2016, pedrozo-penafiel_entanglement_2020, macri_loschmidt_2016}.

\begin{figure}[ht!]
    \centering
    \includegraphics[width=\columnwidth]{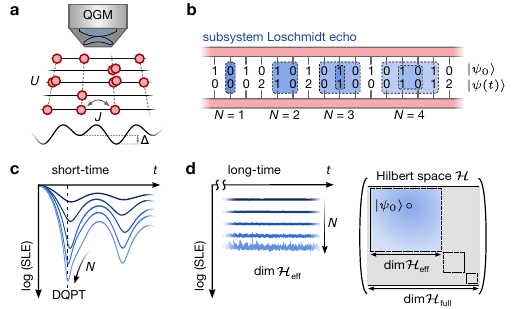}
    \caption{\textbf{Subsystem Loschmidt echo (SLE) and schematic of the experiment.} \textbf{a,} Schematic of the quantum gas microscope (QGM) resolving dynamics in the 1D Bose-Hubbard model with tunnel coupling $J$, on-site interaction $U$, and staggered potential $\Delta$. \textbf{b,} Illustration of density snapshots for the initial state $\ket{\psi_0}=\ket{\dots 101010 \dots }$ and time-evolved state $\ket{\psi(t)}$. The SLE measures the probability of finding the initial state bitstring of length $N$ in the time-evolved state snapshots. \textbf{c, d,} Short- and long-time dynamics of the SLE. \textbf{c,} In the short-time regime, the SLE can exhibit emergent non-analytic features (dashed line) that signal a dynamical quantum phase transition (DQPT). \textbf{d,} In the long-time limit, the scaling of the time-averaged SLE with $N$ quantifies the dimension of accessible Hilbert space ($\mathrm{dim}\,{\mathcal{H}_{\mathrm{eff}}}$), which is reduced compared to the full Hilbert space $\mathcal{H}$ in the presence of kinetic constraints, as illustrated on the right (blue square).}
    \label{fig:sketch}
\end{figure}

Despite its numerous applications, only few studies have explored the LE experimentally. The primary obstacle lies in the fact that the LE is typically exponentially small in system size, and measuring it for even modest one-dimensional (1D) systems accessible to current experiments becomes practically infeasible for system sizes significantly larger than $\sim 10$ sites. All experiments up to date therefore focused on rather small systems~\cite{jurcevic_direct_2017, martinez_real-time_2016, zhu_probing_2024, braumuller_probing_2022}, or special situations where the overlap between the initial and time-evolved state remains high~\cite{cao_interaction-driven_2022, singh_quantifying_2019, guo_observation_2021}.

Motivated by the concept of LE, here we study a quasi-local observable, the \textit{subsystem Loschmidt echo} (SLE), by analysing site-resolved density snapshots in a quantum gas microscope (Fig.~\ref{fig:sketch}a). The SLE quantifies the probability for a subsystem of size $N$ to return back to its initial state ($N\ll L$, where $L$ is the system size)~\cite{bandyopadhyay_observing_2021,halimeh_local_2021, zhang_many-body_2023}. For translationally invariant product initial states, we can define the SLE as a spatial average over local SLEs, ${\mathcal{L}_N^{(i)} = \expect{\prod_i^{i+N-1} \hat{P}_i}}$, where $i$ is the site index and $\hat{P}_i$ is a local projection operator onto the initial state. Pictorially, this procedure is similar to locally comparing two ‘DNA sequences’ corresponding to initial and time-evolved states and measuring the probability of overlaps of strings of a finite length $N$ (Fig.~\ref{fig:sketch}b).

Not only does the SLE capture key properties of the full LE, but it also offers several distinct advantages. While the SLE is exponentially small in $N$, we demonstrate that already small subsystem sizes allow us to draw conclusions about the thermodynamic limit. This is in sharp contrast to the full LE, which is only defined for finite systems and can thus suffer from strong finite-size effects. Additionally, since the SLE is defined locally, it paves the way towards spatially-resolved studies of non-equilibrium dynamics. Lastly, the SLE is robust against experimental noise, such as particle loss, exhibits suppressed temporal fluctuations that complicate measurements of the full LE, and exhibits significantly shorter relaxation times.

In this work, we experimentally probe both the short- and long-time dynamics of the SLE in the 1D Bose-Hubbard model (BHM) with tunable on-site interactions $U$ and tunable staggered potential $\Delta$ (Fig.~\ref{fig:sketch}a), using a caesium quantum gas microscope~\cite{impertro_unsupervised_2023, impertro_local_2024}. The dynamics is governed by the Hamiltonian
\begin{equation}
    \begin{aligned}
        \hat H &= - J \sum_{i} \left(  \hat a_{i}^{\dagger} \hat a_{i+1}^{\phantom{\ast}} + \mathrm{h.c.} \right)\\ & + \frac{U}{2} \sum_{i} \hat n_{i} (\hat n_{i}-1) + \frac{\Delta}{2} \sum_{i} (-1)^i \hat n_{i},
    \end{aligned}
    \label{eq:hamiltonian}
\end{equation}
where $\hat a_{i}^{\dagger}$ and $\hat a_{i}$ are the bosonic creation and annihilation operators, $\hat n_{i}$ is the number operator, and $J$ is the tunnel coupling. Using a Feshbach resonance, $U/J$ can be tuned from non-interacting to $U/J \sim 30$.

This simple model exhibits remarkably rich properties, enabling the exploration of a wide range of out-of-equilibrium dynamics spanning ergodic, integrable and fragmented regimes~\cite{moudgalya_quantum_2022}. In the short-time regime, we study the appearance of a dynamical quantum phase transition (DQPT)~\cite{heyl_dynamical_2013,heyl_dynamical_2018} following a far-from-equilibrium quench in an integrable model ($U\gg J$, $\Delta=0$), and use the SLE to identify its origin in terms of genuine higher-order correlation functions (Fig.~\ref{fig:sketch}c). In the long-time regime, the time-average of the SLE provides quantitative access to the effective dimension of the accessible Hilbert space (Fig.~\ref{fig:sketch}d). By introducing a strong staggered potential $\Delta$ and a resonant on-site interaction $U\approx \Delta$, we realize emergent kinetic constraints that, in the limit of large $\Delta$, suppress all but correlated hopping processes. By measuring the scaling of the time-averaged SLE with subsystem size $N$, we directly reveal the fragmented structure of the Hilbert space and extract the dimension of two distinct fragments in the thermodynamic limit.
\begin{figure*}[htb!]
\includegraphics[width=\textwidth]{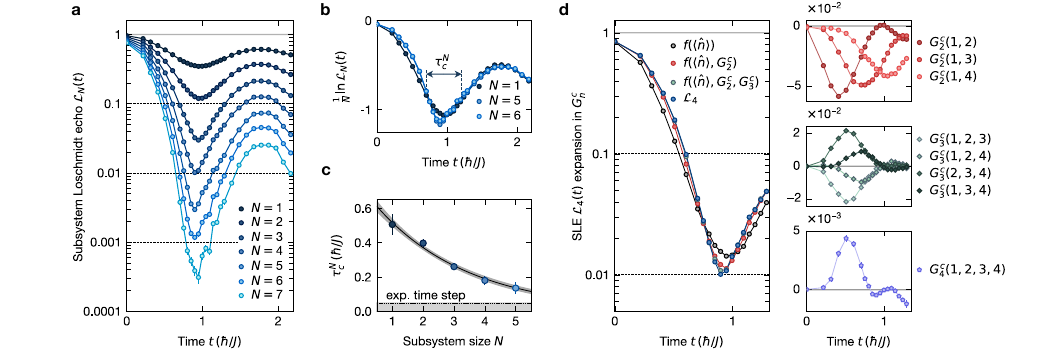}
     \caption{\textbf{Short-time dynamics: dynamical quantum phase transition.} \textbf{a,} SLE evaluated for subsystem sizes $N = 1 - 7$ in the central 32 sites of 40-site long chains to avoid edge effects. A DQPT emerges at $t_c\approx 0.9\hbar/J$ [$J/h=$\SI{155(2)}{Hz}, $\hbar/= h/(2\pi)$ is the reduced Planck's constant]. Error bars are estimated using standard error of proportion and, when not visible, are smaller than the marker. Every data point around the peak was obtained from $\simeq130$ fluorescence images, each containing $\simeq40$ copies of the system. \textbf{b,} Rate function for subsystem sizes $N = 1, 5$ and $6$. \textbf{c,} The characteristic time $\tau_c^N$, extracted using an empirical fit function, decreases exponentially with increasing subsystem size $N$ (see SI for details). The shaded region highlights the time resolution of our experiment. Error bars are the fit errors. Solid line shows an exponential fit to the data, together with the $\pm1\sigma$ confidence interval. \textbf{d,} SLE for subsystem size $N=4$, $\mathcal{L}_4$, together with contributions to the expansion in Eq.~\eqref{eq:subsystemLE_npoint_corrs}, calculated using the experimentally measured higher-order $n$-point connected correlation functions $G_n^c$, shown on the right for $n = 2 - 4$. Error bars show the standard deviation obtained using bootstrapping and if not visible, are smaller than the markers.}
     \label{fig:DQPTs}
\end{figure*}

\begin{center}
\textbf{Short-time dynamics}
\end{center}
\vspace{0.5em}
DQPTs occur after a quench at critical evolution times $t_c$, where the wavefunction overlap between the initial state $\ket{\psi_0}$ and the time-evolved state $\ket{\psi(t_c)}$, quantified by the LE, $\mathcal{L}(t_c) = |\braket{\psi_0|\psi(t_c)}|^2$, nearly vanishes while developing non-analytic, cusp-like features. A general understanding of DQPTs is still lacking, but it is believed that DQPTs appear when the system is taken sufficiently far from equilibrium, e.g., by quenching across an equilibrium critical point~\cite{heyl_dynamical_2013,heyl_dynamical_2018}.

An intriguing interpretation connects DQPTs to large deviation theory, a statistical mechanics framework that deals with probabilities of extremely rare and atypical events~\cite{touchette_large_2009}. In classical stochastic mechanics, detecting a non-analytic feature in the probability distribution of a rare event, such as identifying a low-energy ferromagnetic configuration in a very hot system, can indicate a sudden and dramatic change in the system's history and could even provide insight into the underlying equilibrium critical point~\cite{baek_singularities_2015, meibohm_finite-time_2022}. Similarly, in a quantum system, finding an initial state configuration is an extremely rare event even after a relatively short amount of time. A DQPT can then be interpreted as an effective \textit{wall in time}: before the DQPT, we can always assign an initial state to the time-evolved state via analytic continuation. However, this is no longer possible immediately after the DQPT. Using our DNA analogy, the DQPT effectively scrambles the DNA and erases the information about the ‘ancestral’ initial state.

Indirect signatures of DQPTs have been observed in several experiments, including the dynamical appearance of vortices~\cite{flaschner_observation_2018}, time-averaged two-body correlations and the mean size of the largest domain wall~\cite{zhang_observation_2017}, or the time-averaged magnetization~\cite{xu_probing_2020}. By probing a modified return probability, closely related to the LE, a DQPT was observed in a ten-qubit trapped ion quantum simulator~\cite{jurcevic_direct_2017}. Here, we experimentally demonstrate that DQPTs are captured by the SLE~\cite{bandyopadhyay_observing_2021,halimeh_local_2021}. Moreover, we demonstrate that the DQPT observed directly via the SLE arises from genuine higher-order correlations, identifying it as a non-trivial many-body phenomenon.

We start by preparing a charge density wave (CDW) initial state $\ket{\psi_0} = \ket{\cdots 10101010 \cdots}$ with $L=40$ as the ground state of a half-filled bichromatic optical superlattice with large potential energy offset on every other site (mean occupation $\sim$93\% on the occupied, and $\sim$2\% on the empty sites). Once residual coherences have dephased, we initiate the dynamics by suddenly lowering the primary lattice depth. After the quench the system evolves in the integrable regime of the BHM [Eq.~(\ref{eq:hamiltonian})] with hard-core interactions and no staggered potential ($U/J = 25.4(4)$, $\Delta = 0$). Evaluating the spatially-averaged SLE as a function of time for different subsystem sizes $N = 1 - 7$ reveals an increasingly sharp feature for increasing $N$ around $\approx0.9$ tunnelling times (Fig.~\ref{fig:DQPTs}a). This steepening and emergent kink-like behaviour in the SLE is a defining feature of a DQPT~\cite{bandyopadhyay_observing_2021, halimeh_local_2021}.

In analogy to equilibrium phase transitions, the system's response in non-equilibrium situations can be analysed via the time-evolution of the \textit{dynamical free energy density}, or the \textit{rate function}, defined as $-\mathrm{ln}\,\mathcal{L}_N/N$~\cite{heyl_dynamical_2018}. As expected, the rate function evaluated using the measured SLE in Fig.~\ref{fig:DQPTs}a reveals a dynamical critical point (Fig.~\ref{fig:DQPTs}b). Moreover, we find that except for very small subsystems ($N=1$) the data for larger subsystems collapse except at the critical point, where residual deviations between $N=5$ and $N=6$ are noticeable. Indeed, the rate function is expected to become approximately constant with increasing $N$ if the subsystem is sufficiently large to capture all relevant correlations in the system, which suggests that correlations might play an important role at the critical point, as we explain below.

In the thermodynamic limit, an ideal DQPT observed via the full LE is characterized by a piecewise linear function. In Fig.~\ref{fig:DQPTs}c, we provide a more in-depth analysis of the sharpening of the rate function $\mathrm{ln}\,\mathcal{L}_N/N$ as a function of the subsystem size. Using an empirical fit function, we extract a characteristic time, $\tau_c^N$, during which the experimental signal deviates from a piecewise linear function with $\tau_c = 0$. Our results show that the characteristic time $\tau_c^{N}$ approaches zero with increasing subsystem size, consistent with an exponential scaling, hence, providing further evidence that the SLE reliably captures key properties of the LE.

Intriguingly, the SLE allows to gain additional insight into the origin of the DQPT by noting that it can be expressed in terms of higher-order connected correlation functions (see SI for details)
\begin{equation}
    \begin{aligned}
        \mathcal{L}_N = f\bigl( \langle \hat{n}_i\rangle, G_2^c\left( 
i_1, i_2\right), \cdots, G_N^c\left( 
i_1, i_2, \dots i_N \right) \bigr),
    \end{aligned}
    \label{eq:subsystemLE_npoint_corrs}
\end{equation}
where e.g. $G_2^c(i,j) = \langle \hat{n}_i \hat{n}_j \rangle - \langle \hat{n}_i \rangle\langle \hat{n}_j \rangle$ is the two-point connected correlation function. For instance, for the CDW initial state, the SLE for $N=1$ can be expressed only using local mean fillings $n_e, n_o$ on even and odd sites, respectively, ${\mathcal{L}_1 = (n_e+1-n_o)/2}$, but for $N=2$, 2-point correlations have to be included to accurately reproduce the SLE, ${\mathcal{L}_2 = n_e(1-n_o) - G_2^c(i,i+1)}$. Each higher-order $n$-point connected correlation function $G_n^c\left(i_1, i_2, \cdots, i_n\right)$ contains distinct information that cannot be captured by lower-order correlations alone. Correspondingly, the measurement of higher-order correlation functions in previous experiments allowed estimating the level of complexity of a many-body state~\cite{zheng_efficiently_2022, schweigler_experimental_2017, hodgman_direct_2011, dall_ideal_2013, rispoli_quantum_2019, koepsell_microscopic_2021}.

\begin{figure*}[htb!]
\includegraphics[width=\textwidth]{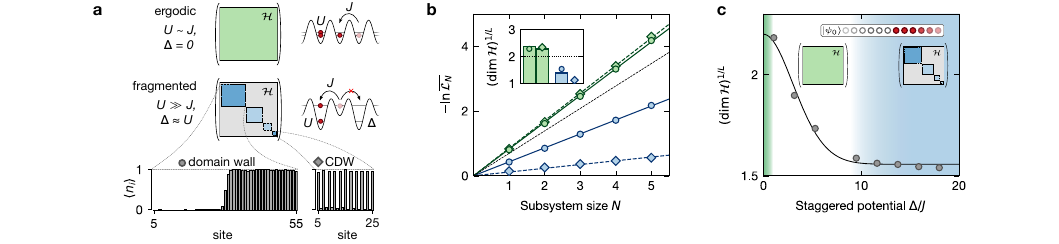}
     \caption{\textbf{Long-time dynamics: structure and dimension of the many-body Hilbert space.} \textbf{a,} The staggered 1D BHM in ergodic and fragmented regimes. In the ergodic regime, both the CDW and domain wall initial states explore the entire Hilbert space. For strong kinetic constraints, the two initial states live in different fragments of the Hilbert space. \textbf{b,} Time-averaged $-\mathrm{ln}\,\overline{\mathcal{L}_N}$ as a function of the subsystem size $N$. We take the spatial average of $\mathrm{ln}\,\overline{\mathcal{L}_N^{(i)}}$, where $i$ is the spatial index of a subsystem. The slope (inset) directly measures the dimension of accessible Hilbert space. The dashed black line illustrates $\left( \mathrm{dim}\,\mathcal{H}_{\mathrm{eff}} \right)^{1/L} = 2$. The markers are experimental data, and the bars (in the inset) are numerical predictions. For the CDW initial state, central $28$ sites of the total $30$ sites were used for the analysis. For the domain wall initial state, central $52$ sites of the total $60$ sites were used. In the ergodic regime, the analysis region for the domain wall was further reduced to central $20$ sites. We note that we experimentally prepare the CDW initial state with atoms occupying the high-energy lattice sites. The error bars in the main panel show the standard deviation obtained using bootstrapping (fit errors in the inset). If not visible, they are smaller than the markers. \textbf{c,} Dimension of the accessible Hilbert space for a domain wall initial state as a function of varying strength of the staggered potential $\Delta/J$, keeping $\Delta\approx U$. The solid line shows a Gaussian fit as a guide to the eye. The error bars (fit errors) are smaller than the markers.}
     \label{fig:IPR}
\end{figure*}

In Fig.~\ref{fig:DQPTs}d, we show the experimentally measured higher-order $n$-point connected correlation functions ($n = 2 - 4$) and plot their relative contribution to the SLE $\mathcal{L}_4$. Interestingly, we reveal that at the dynamical critical point, where the SLE is the smallest, the $3$- and $4$-point correlations play an important role in the sharpening of the signal and the emergence of the non-analytic feature. For longer strings, the SLE - or indeed the LE - is exponentially smaller and therefore exponentially more sensitive to relative contributions from increasingly smaller higher-order correlations. Our observation, therefore, strongly suggests that the observed DQPT is a non-trivial many-body phenomenon. The situation is reminiscent of emergence of \textit{equilibrium} phase transitions, where non-analyticities in the thermodynamic free energy arise due to high-order correlation functions~\cite{kardar_statistical_2007, rispoli_quantum_2019}.

It is also quite striking that the relative contribution from the higher-order correlations is important already in such early dynamics of the many-body system. Indeed, for hard-core bosons in 1D (i.e. free fermions), the Lieb-Robinson theorem bounds the speed of information spreading to $4J/\hbar$~\cite{paredes_tonksgirardeau_2004, cheneau_light-cone-like_2012, wienand_emergence_2024}. In accordance, we see that the higher-order correlations are exponentially small for $t\lesssim \hbar/J$, since they must originate from the exponential tail beyond the Lieb-Robinson cone. However, the SLE at the dynamical critical point is still highly sensitive to these small contributions.

Our observation regarding the relative importance of higher-order correlations in the emergence of DQPT is likely general, as low-order correlations, being local operators, must be analytic for finite times following the quench~\cite{avdoshkin_euclidean_2020}. Therefore, the non-analyticity in the LE must arise from considering all higher-order correlations.

\begin{center}
\textbf{Long-time dynamics}
\end{center}
\vspace{0.5em}
Having demonstrated that the SLE accurately captures key features of the LE in the short-time dynamics, we now turn to experiments probing its long-time dynamics. Our key insight is that starting with a high-energy initial state, the long-time average of the SLE in quasi-equilibrium, $\overline{\mathcal{L}_N}$, contains information about the effective thermodynamic entropy of the system, ${- \mathrm{ln}\,\overline{\mathcal{L}_N} \equiv S_N^{\textrm{eff}}}$, and thus the dimension of the effective \textit{accessible} Hilbert space, $\mathrm{dim}\,\mathcal{H}_{\mathrm{eff}}$ (Fig.~\ref{fig:sketch}d). Intuitively, this is because for thermalizing high-energy initial states, all eigenstates within the accessible Hilbert space will typically be roughly equally populated and therefore the probability of finding a particular string will be inversely proportional to $\mathrm{dim}\,\mathcal{H}_{\mathrm{eff}}$.

More precisely, we find that for subsystems of size $N$, larger than the correlation length of the time-evolved state $\xi$ in our measurement basis, the effective entropy is upper-bounded by the thermodynamic entropy $S_N(\beta,\mu)$, since
\begin{equation}
    \begin{aligned}
    -\mathrm{ln}\, \overline{\mathcal{L}_N} &= S_N(\beta,\mu) - \frac{1}{2}\beta^2 \delta E^2 + \mathcal{O}(\beta^3 \delta E ^3) \\&\leq S_N(\beta,\mu).
    \end{aligned}
\end{equation}
Here, $\beta$ is the inverse temperature, $\mu$ is the chemical potential and $\delta E^2$ is the energy variance of the eigenstates of the post-quench Hamiltonian. For high-temperature initial states, $\beta \to 0$, and thus the SLE directly measures the thermodynamic entropy $-\mathrm{ln}\, \overline{\mathcal{L}_N} \approx S_N(\beta,\mu)$. Importantly, the entropy $S_N(\beta,\mu)$ corresponds to the one expected for an infinitely large system, i.e., in the thermodynamic limit. This is easily understood by adopting a picture where the large system acts as a grand-canonical bath for the subsystem (see SI).

Extrapolating to the thermodynamic limit ($L \to \infty$), we can relate the effective thermodynamic entropy, which we assume to be extensive in $N$, to the dimension of the accessible Hilbert space as

\begin{equation}
    \begin{aligned}
        \left( \mathrm{dim}\,\mathcal{H}_{\mathrm{eff}} \right)^{1/L} \approx \mathrm{exp}\left( \diff{\left(-\mathrm{ln}\,\overline{\mathcal{L}_N}\right)}{N} \right).
    \end{aligned}
    \label{eq:slope_vs_dimH}
\end{equation}

We note that our observation about the SLE mirrors the fact that the time-averaged full LE is exactly equal to the R\'enyi entropy of the diagonal ensemble, i.e., the inverse participation ratio $\overline{\mathcal{L}} \equiv \mathrm{IPR} = \sum_n |\braket{n|\psi_0}|^4$, where $\ket{n}$ are eigenstates of the post-quench Hamiltonian~\cite{torres-herrera_dynamics_2015,karamlou_quantum_2022, guo_observation_2021}. For equally populated eigenstates, the IPR is indeed also a measure of the dimension of the accessible Hilbert space. Note, however, that the R\'enyi entropy, that we would extract from the LE, and the thermodynamic entropy we extract using the SLE are different, since the full LE does not satisfy ensemble equivalence.

Here, we measure the long-time average of the SLE in the ergodic regime and in the presence of strong kinetic constraints, that give rise to \textit{Hilbert space fragmentation} (HSF) -- a phenomenon where a system can avoid thermalization and violate the eigenstate thermalization hypothesis despite the absence of disorder~\cite{dalessio_quantum_2016}. HSF is characterized by a decoupling of the many-body Hilbert space into numerous (approximately) independent blocks, known as \textit{fragments}~\cite{khemani_localization_2020, pai_localization_2019, sala_ergodicity_2020, moudgalya_quantum_2022} (Fig.~\ref{fig:IPR}a). The block structure arises due to strong kinetic constraints rather than global symmetries of the Hamiltonian, such as particle number or energy. An intriguing signature of HSF is that initial states that are prepared in different fragments exhibit very different dynamics, e.g., certain initial states may be completely frozen, while others thermalize, however, in a restricted subspace. First signatures of this behaviour have been explored experimentally with ultracold atoms in tilted one- and two-dimensional optical lattices by probing the initial state memory for different product initial states~\cite{kohlert_exploring_2023, scherg_observing_2021,adler_observation_2024}. While this gives a first indication of HSF, this analysis does not reveal information about the dimension of the respective fragment. The SLE, however, gives direct experimental access to the intricate structure of the Hilbert space for fragmented models.

In the experiment, we probe the long-time dynamics in the 1D staggered BHM [Eq.~\eqref{eq:hamiltonian} and Fig.~\ref{fig:IPR}a]. For $U/J \sim 1$ and $\Delta = 0$, we realize an ergodic model. In the resonant regime, where $\Delta \approx U$, increasing $\Delta/J$ suppresses all but correlated hopping processes, leading to emergence of kinetic constraints. In the regime where $\Delta/J$ is very large, we can derive a perturbative first-order effective Hamiltonian
\begin{equation}
    \begin{aligned}
        \hat{H}_{\text{frag}} &= -J \left( \sum_{\text{odd}\,i} \left( |11\rangle_{i,i+1}\langle 20|_{i,i+1} + \text{h.c.} \right) \right. \\
        &\quad \left. + \sum_{\text{even}\,i} \left( |11\rangle_{i,i+1}\langle 02|_{i,i+1} + \text{h.c.} \right) \right).
    \end{aligned}
    \label{eq:h_eff}
\end{equation}
This Hamiltonian supports strong HSF, allowing us to experimentally explore the thermalization and \textit{quasi-equilibration} under HSF at intermediate timescales. Eventually, higher-order terms in the perturbative description couple the fragments, and lead to full thermalization of the system.

We probe the dynamics starting with CDW and domain-wall initial states of the form $|\cdots 00001111\cdots\rangle$. Under the effective approximate Hamiltonian, the CDW initial state will be frozen (corresponding to an accessible Hilbert space of dimension of 1), while a single-domain wall initial state will exhibit strong dynamical evolution. In the ergodic regime, we set $U/J=2.7(1)$ and $\Delta = 0$. In the fragmented regime, we set $U/J=16.8(2)$ and $\Delta \approx U$. The tunnel coupling for all experiments is set to $J/h=\SI{122(1)}{Hz}$. To resolve double occupancies and avoid parity projection, we prepare the initial state only in odd chains, leaving the even chains empty. After performing the quench experiment, we freeze the on-site occupations and use the neighbouring empty chains to split the doublons, thereby avoiding parity projection in reading out the original on-site occupations (we estimate that the triple occupancies are very rare).

In Fig.~\ref{fig:IPR}b we show the SLE averaged over four measurements taken at $t\approx(230 - 280)\hbar/J$ after the quench, sufficiently long to assume that the system has reached quasi-equilibrium. Interestingly, we observe a very good linear scaling between the time-averaged dynamical free energy, $-\mathrm{ln}\,\overline{\mathcal{L}_N}$, and the subsystem size $N$ already for $N \sim 1$. This means that the correlation length of the quasi-equilibrated high-temperature state is very small, $\xi\sim 1$. The situation might be different for states that relax to a finite temperature with a stronger role of correlations. For such states, it might be necessary to probe larger subsystem sizes $N$ to observe the regime where the rate function $-\mathrm{ln}\,\overline{\mathcal{L}_N}/N$ is constant.

We can now use the slope of the data in Fig.~\ref{fig:IPR}b to directly quantify the dimension of the accessible Hilbert space $\left(\mathrm{dim}\,\mathcal{H}_{\mathrm{eff}}\right)^{1/L}$ in the thermodynamic limit $L \to \infty$ [Eq.~\eqref{eq:slope_vs_dimH}]. The inset in Fig.~\ref{fig:IPR}b shows the measured $\left( \mathrm{dim}\,\mathcal{H}_{\mathrm{eff}} \right)^{1/L}$. In the ergodic regime, we obtain $2.29(1)$ for the case of the domain wall and $2.34(1)$ for the CDW case, consistent with theoretical predictions $2.39$ and $2.28$ respectively (obtained using a Krylov subspace method simulation of the full time-dynamics of the SLE). For comparison, for a high-temperature uncorrelated spin-1/2 system, we would expect $\left( \mathrm{dim}\,\mathcal{H}_{\mathrm{eff}} \right)^{1/L} = 2$, while for a fragment with completely frozen dynamics $\left( \mathrm{dim}\,\mathcal{H}_{\mathrm{eff}} \right)^{1/L} = 1$. Similarly, based on simple combinatorics and by calculating the \textit{von Neumann entropy} in the grand canonical ensemble, for non-interacting bosons at half-filling we expect $\left( \mathrm{dim}\,\mathcal{H}_{\mathrm{eff}} \right)^{1/L} \approx 2.60$ and $2.46$ if we restrict the occupation to at most two particles (see SI for details).

The experimentally measured values are almost identical for the two initial states, consistent with the notion of ergodicity. We note that the measured $\left( \mathrm{dim}\,\mathcal{H}_{\mathrm{eff}} \right)^{1/L}$ for finite interactions is smaller than the one for non-interacting bosons but closer to our prediction allowing only up to two atoms per lattice site. This can be understood by noting that despite the high energy of the initial state, at finite interactions the high-energy states, such as those with many triply- and multiply-occupied lattice sites, will not be accessible, and the temperature will be high but not infinite.

In the fragmented regime, we find a very different result of $\left( \mathrm{dim}\,\mathcal{H}_{\mathrm{eff}} \right)^{1/L} = 1.54(1)$ and $1.12(1)$ for the domain wall and CDW respectively. From numerical (time-dependent variational principle) simulations we expect $1.39$ and $1.01$. The accessible Hilbert space dimension differs significantly between the two initial states, directly showing that they indeed live in different uncoupled fragments with different sizes. The residual deviation between theory and experiment in the fragmented regime can be explained by imperfect initial state preparation and small atom loss during the dynamics. We note that the numerical prediction is in an excellent agreement with a simple combinatorial calculation based on the effective Hamiltonian, that yields dimensions $\sqrt{2}$ and $1$ for domain wall and CDW initial states respectively (see SI).

Lastly, we demonstrate the usefulness and simplicity of our new observable by probing the emergence of HSF with increasing strength of the staggered potential $\Delta/J$, keeping $\Delta \approx U$. In previous work on tilted 1D Fermi-Hubbard chains, it has been noticed that the HSF is robust even far away from the perturbative limit of $\Delta/J \to \infty$~\cite{scherg_observing_2021}. Using standard measures, such as the imbalance, it is however rather difficult to make conclusive statements about the emergence of HSF and the cross-over regime, where higher-order terms in the effective Hamiltonian play a significant role.

Here, we simply perform the experiment for multiple values of $\Delta/J$ and measure the dimension of the accessible Hilbert space (see Fig.~\ref{fig:IPR}c). The time window for the measurement is again approximately $(230 - 280) \hbar/J$. While for small $\Delta/J$, we find $\left( \mathrm{dim}\,\mathcal{H}_{\mathrm{eff}} \right)^{1/L}$ of an ergodic system, for $\Delta/J \gtrsim 10$ we directly reveal the presence of the kinetic constraints that restrict the accessible Hilbert space. We note that by choosing a later time-window, the cross-over regime would move to higher $\Delta/J$, as the higher-order terms in the effective Hamiltonian ultimately always lead to a full thermalization.

In summary, in this work we have introduced the SLE as a powerful and simple new quantity to analyse quantum dynamics in many-body systems and provided direct experimental proof of its usefulness. Our approach is applicable to a broad range of quantum simulation and computing platforms and can be further extended to other initial states or different protocols, where the systems are expected to thermalize at a lower temperature. Measuring the SLE would then allow extracting the entropy and the effective Hilbert space dimension of the system at different energy densities. The LE, and thus the SLE, is also deeply connected to typical dynamical probes of quantum chaos, such as the spectral autocorrelation function and, therefore, contains information about energy-level correlations~\cite{torres-herrera_dynamical_2017-1, spectral_correlation}. In future work, it might be interesting to extend the SLE measurement beyond product initial states.

Our results are also naturally extendable to large two-dimensional systems~\cite{manetsch_tweezer_2024, impertro_unsupervised_2023, kwon_site-resolved_2022, tao_high-fidelity_2024}. Exploring the SLE in these settings promises valuable insights into non-equilibrium dynamics, quantum chaos~\cite{peres_stability_1984, gorin_dynamics_2006, haake_quantum_2018, torres-herrera_dynamical_2017-1, spectral_correlation}, false vacuum decay in 1D and 2D lattice gauge theories~\cite{schwinger_gauge_1951, cohen_schwinger_2008}, and quantum metrology~\cite{li_improving_2023, colombo_time-reversal-based_2022, davis_approaching_2016, pedrozo-penafiel_entanglement_2020, macri_loschmidt_2016}. Measuring the thermodynamic entropy in a space- and time-resolved manner could provide valuable insight into, for example, the intricate dynamical evolution in many-body localized systems~\cite{abanin_colloquium_2019}. Moreover, we expect that the SLE can effectively identify and probe thermalization dynamics in more exotic regimes, such as long-range or frustrated models, where the entropy can be sub-extensive. Furthermore, the string observables we examine display intriguing dynamics characterized by anomalously large temporal fluctuations, making them a compelling subject for future experimental and theoretical investigation.

\vspace{2em}
\begin{center}
\textbf{ACKNOWLEDGEMENTS}
\end{center}
\vspace{0.5em}

The authors would like to acknowledge insightful discussions with Sarang Gopalakrishnan, Maksym Serbyn and Norman Yao. We would also like to thank Emma Cussenot for early contributions to the project and Eliska Simsova for valuable feedback on the manuscript. We received funding from the Deutsche Forschungsgemeinschaft (DFG, German Research Foundation) via Research Unit FOR5522 under project number 499180199 and under Germany’s Excellence Strategy – EXC-2111 – 390814868 and from the German Federal Ministry of Education and Research via the funding program quantum technologies – from basic research to market (contract number 13N15895 FermiQP). This publication has further received funding under Horizon Europe programme HORIZON-CL4-2022-QUANTUM-02-SGA via the project 101113690 (PASQuanS2.1).
S.H. was supported by the education and training program of the Quantum Information Research Support Center, funded through the National research foundation of Korea (NRF) by the Ministry of science and ICT (MSIT) of the Korean government (No.2021M3H3A1036573). I.P.R. received funding from the International Max Planck Research School (IMPRS) for Quantum Science and Technology. Z.-H. S. and M.H. received funding from the European Research Council (ERC) under the European Union’s Horizon 2020 research and innovation programme (grant agreement No. 853443). S.B. and A.P. were supported by NSF Grant DMR-2103658 and the AFOSR Grant FA9550-21-1-0342. This research was supported in part by grant NSF PHY-2309135 to the Kavli Institute for Theoretical Physics (KITP).

\vspace{2em}
\begin{center}
\textbf{REFERENCES}
\end{center}
\vspace{0.5em}

\putbib[manuscript]
\end{bibunit}
\clearpage

%% file: source/title.tex
\title{Probing quantum many-body dynamics using subsystem Loschmidt echos}

\author{Simon~Karch}\email{Simon.Karch@physik.uni-muenchen.de}
    \affiliation{Fakult\"{a}t f\"{u}r Physik, Ludwig-Maximilians-Universit\"{a}t, 80799 Munich, Germany}
    \affiliation{Max-Planck-Institut f\"{u}r Quantenoptik, 85748 Garching, Germany}
    \affiliation{Munich Center for Quantum Science and Technology (MCQST), 80799 Munich, Germany}
\author{Souvik~Bandyopadhyay}
    \affiliation{Department of Physics, Boston University, 590 Commonwealth Ave., Boston, MA 02215, USA}
\author{Zheng-Hang~Sun}
    \affiliation{Theoretical Physics III, Center for Electronic Correlations and Magnetism, Institute of Physics, University of Augsburg, D-86135 Augsburg, Germany}
\author{Alexander~Impertro}
\author{SeungJung~Huh}
\author{Irene~Prieto~Rodr\'iguez}
\author{Julian~F.~Wienand}
    \affiliation{Fakult\"{a}t f\"{u}r Physik, Ludwig-Maximilians-Universit\"{a}t, 80799 Munich, Germany}
    \affiliation{Max-Planck-Institut f\"{u}r Quantenoptik, 85748 Garching, Germany}
    \affiliation{Munich Center for Quantum Science and Technology (MCQST), 80799 Munich, Germany}
\author{Wolfgang~Ketterle}
    \affiliation{Department of Physics, Massachusetts Institute of Technology, Cambridge, MA 02139, USA}
    \affiliation{MIT-Harvard Center for Ultracold Atoms, Cambridge, MA 02139, USA}
\author{Markus~Heyl}
    \affiliation{Theoretical Physics III, Center for Electronic Correlations and Magnetism, Institute of Physics, University of Augsburg, D-86135 Augsburg, Germany}
    \affiliation{Centre for Advanced Analytics and Predictive Sciences (CAAPS), University of Augsburg, Universitätsstr. 12a, 86159 Augsburg, Germany}
\author{Anatoli~Polkovnikov}
    \affiliation{Department of Physics, Boston University, 590 Commonwealth Ave., Boston, MA 02215, USA}
\author{Immanuel~Bloch}
\author{Monika~Aidelsburger}\email{Monika.Aidelsburger@physik.uni-muenchen.de}
    \affiliation{Fakult\"{a}t f\"{u}r Physik, Ludwig-Maximilians-Universit\"{a}t, 80799 Munich, Germany}
    \affiliation{Max-Planck-Institut f\"{u}r Quantenoptik, 85748 Garching, Germany}
    \affiliation{Munich Center for Quantum Science and Technology (MCQST), 80799 Munich, Germany}

%% file: source/supplements.tex
\begin{bibunit}
\setcounter{section}{0}
\setcounter{equation}{0}
\setcounter{figure}{0}
\setcounter{table}{0}
\renewcommand{\theequation}{S\arabic{equation}}
\renewcommand{\theHequation}{S\arabic{equation}}
\renewcommand{\thefigure}{S\arabic{figure}}
\renewcommand{\theHfigure}{S\arabic{figure}}
\renewcommand{\thetable}{S\arabic{table}}
\renewcommand{\theHtable}{S\arabic{table}}
\setcounter{page}{1}

\title{Supplementary Information for: \\ Probing quantum many-body dynamics using subsystem Loschmidt echos}

\maketitle
\tableofcontents


\section{Experimental details}

\subsection{Experimental sequence}

\begin{figure*}
    \centering
    \includegraphics[]{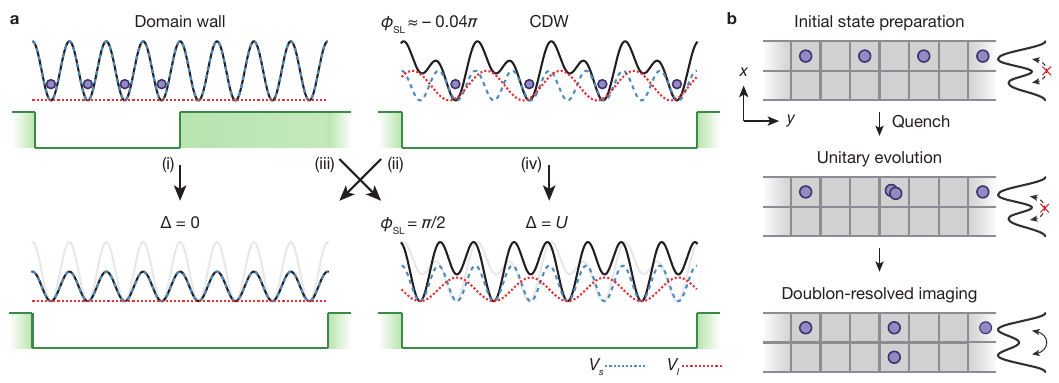}
    \caption{\textbf{Illustration of the experimental sequence.} \textbf{a,} The two initial states -- domain wall and charge density wave (CDW) -- are prepared using a 2D superlattice and a programmable repulsive box potential (green), generated using a digital micromirror device (DMD). The domain wall state is prepared by introducing an energy offset on half of the sites along the $y$-axis, while the CDW state is prepared by applying energy offset using a tilted superlattice potential along the $y$-axis. After preparing the initial state, we switch the DMD potential to a simple box potential and change the phase of the superlattice, $\phi_{\mathrm{SL}}$, to realize a staggered configuration, keeping the initial state frozen. Lastly, the dynamics is initiated by suddenly quenching the short-lattice depth. \textbf{b,} For the long-time measurement, we prepare the initial state only in every other chain along the $x$-axis, keeping half of the chains empty. This is achieved by adding an additional energy offset on every other site along the $x$-axis using a tilted superlattice potential. After performing the quench experiment, we freeze the density distribution of the atoms by suddenly increasing the short lattice depth along the $y$-axis. Before taking the fluorescence image of the density distribution, we first split the doubly-occupied sites along the $x$-axis using a double-well potential.}
    \label{fig:exp_seq}
\end{figure*}

In this section, we describe the experimental sequences used for initial-state preparation and quench experiments.

As described in Refs.~\cite{impertro_unsupervised_2023, wienand_emergence_2024,impertro_local_2024, impertro_realization_2024}, our experiments start by preparing a Bose-Einstein condensate of $^{133}\rm{Cs}$ atoms in a single plane of a vertical lattice with $\SI{8}{\micro\m}$ spacing (large-spacing vertical lattice). Within the plane, the atoms are confined in a repulsive box potential projected using a digital micromirror device (DMD), illuminated with a blue-detuned, incoherent light from a multi-mode laser diode with a central wavelength of $\lambda = 525\,$nm. We then perform additional evaporative cooling by lowering the vertical lattice depth. To avoid doublon formation during the lattice ramp into the Mott insulating phase with one atom per site, the box height is set to be $\sim U/2$. This causes excess atoms inside the box to spill over, enabling preparation of high-quality initial states such as unity-filling Mott insulator, charge-density wave, or the long-lattice Mott insulator (with filling $1/4$ in the primary short lattice). Further details about the experimental setup can be found in our previous work~\cite{impertro_unsupervised_2023, wienand_emergence_2024,impertro_local_2024, impertro_realization_2024}.

The superlattice potential realized in the experiment can be expressed as
\begin{equation}
    V(x) = V_\mathrm{s} \cos^2\left(\pi x/a_\mathrm{s}\right) + V_\mathrm{l} \cos^2\left(\pi x /a_\mathrm{l} + \phi_{\mathrm{SL}}/2\right),
\end{equation}
where $V_\mathrm{s(l)}$ is the potential depth of the short(long)-period lattice, and $\phi_{\mathrm{SL}}$ is the superlattice phase, controlled using a frequency-offset lock~\cite{impertro_local_2024}.

\textbf{Short-time dynamics (DQPT):} In order to study the short-time dynamics of the SLE, we start by preparing a charge-density wave (CDW) initial state as the ground state of a bichromatic optical superlattice along the $y-$axis with energy offset on every other site of the chain. To this end, we adiabatically ramp up the long lattice along the $y$-axis to 60$E_\mathrm{r,l}$ as well as the short lattice along the $x$-axis to 45$E_\mathrm{r,s}$. Here, $E_\mathrm{r,s(l)} = h^2/2ma_\mathrm{s(l)}^2$ is the recoil energy of short (long) lattice, where $h$ is Planck's constant, $m$ is the atomic mass of Cs, and $a_\mathrm{s} = 383.5\,$nm $(a_\mathrm{l} = 2a_s)$ is the short (long) lattice constant. Simultaneously with the lattice ramp, we increase the magnetic field from \SI{22.4}{G} to \SI{30.2}{G} to increase the scattering length from $270\,a_0$ to $540\,a_0$. Subsequently, we ramp up the short lattice along the $y$-axis to $60E_\mathrm{r,s}$, such that all atoms localize in the low-energy sites of the superlattice potential. The typical filling of our initial state is 93$\%$ in the occupied sites and 2$\%$ in the empty sites.

To achieve stronger vertical confinement and, therefore, a higher on-site interaction energy, we transfer the atoms into a vertical lattice with shorter spacing ($\approx \SI{1.06}{\micro\m}$). Before the transfer, we freeze the on-site occupations by ramping up both short lattices to 60$E_\mathrm{r,s}$, and squeeze the wave function along $z$ by ramping up the large-spacing vertical lattice within \SI{20}{ms}. Subsequently, we turn on the short-spacing vertical lattice within \SI{100}{ms}, thereby increasing the trap frequency to $2\pi \times \SI{6.6}{kHz}$. We further control the relative phase between the short- and long-spacing vertical lattice to ensure that the atoms are loaded into a single layer only. Lastly, the large-spacing vertical lattice is adiabatically switched off within \SI{200}{ms}.

To initiate the far-from equilibrium quench dynamics, we rapidly (within \SI{0.15}{ms}) decrease the short lattice depth along the $y$-axis to $5.3\,E_\mathrm{r,s}$, setting the tunnel coupling $J/h$ to $\SI{155(2)}{Hz}$. At these lattice parameters, and a magnetic field of \SI{30.2}{G}, we obtain an on-site interaction energy of around $U/J \approx 25$, i.e., deep in the hard-core regime.

\textbf{Long-time dynamics [$\dimH$]:} In Fig.~\ref{fig:exp_seq}, we illustrate the experimental sequence used to probe the long-time dynamics. To avoid parity projection of doubly-occupied sites during the fluorescence imaging, we prepare the initial state only in odd chains along the $x$-axis, keeping the even chains empty for subsequent doublon-resolved imaging. In the occupied chains, we initialize the system in one of two different product states: domain wall or CDW. The short-spacing vertical lattice has a trap frequency of $2\pi \times \SI{5.1}{kHz}$ for the long-time quench experiment.

The domain wall initial state preparation is similar to the CDW preparation described above, but involves an additional repulsive potential projected using the DMD, locally increasing the potential energy in one half of the chain by $\gtrsim U$. Thus, when loading the initial product state, we keep half of the chain along the $y$-axis empty (Fig.~\ref{fig:exp_seq}a, left).

To prepare the CDW in alternating chains along the $x$-axis, we prepare a Mott insulator in the long lattices (i.e., a state with 1/4 filling in the primary short lattices). We ramp the long lattices along both $x$- and $y$-axis to $45\,E_\mathrm{r,l}$ in $\SI{220}{ms}$. Then, we ramp the two short lattices to $45\,E_\mathrm{r,s}$ in $\SI{30}{ms}$ with both superlattice phases set to about $\phi_{\mathrm{SL}}\approx-0.04\pi$, resulting in one atom being localized to the site with the lowest potential energy in the superlattice unit cell of $2 \times 2$ sites. Afterwards the long lattices are removed within $\SI{30}{ms}$ and the short-lattices are ramped up to $60\,E_\mathrm{r,s}$.

Prior to the quench that initializes the dynamics, the atoms are transferred from the large-spacing vertical lattice to the short-spacing vertical lattice using the same sequence as in the DQPT experiment. During the transfer, we maintain the short lattices at a depth of $60\,E_\mathrm{r,s}$ to prevent unwanted tunnelling of atoms.

To initiate the dynamics, we quench the system to a symmetric or staggered lattice configuration at a finite tunnel coupling in $\SI{0.15}{ms}$. For a symmetric lattice configuration, we abruptly ramp the short lattice along the $y-$axis to $6.2\,E_\mathrm{r,s}$, corresponding to the tunnel coupling of $J/h = \SI{122(1)}{Hz}$. To realize a lattice configuration with a staggered potential, before quenching the short lattice depth, we tune the superlattice phase to $\pi/2$ and ramp the long lattice along the $y$-axis to the final value within \SI{60}{ms}. During the lattice ramp, we simultaneously change the magnetic field to tune the on-site interaction energy $U$. In the staggered configuration, the staggered potential $\Delta$ and the on-site interaction $U$ are set to be approximately equal.

\begin{figure*}[htb!]
\includegraphics[bb = 0 0 6.30in 1.97in,width=2\columnwidth]{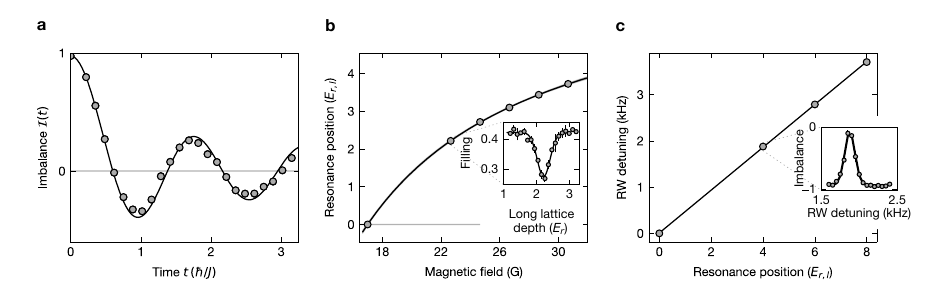}
     \caption{\textbf{Calibration of the tunnel coupling $J$ and the on-site interaction $U$.} \textbf{a,} Calibration of the tunnel coupling $J$ for the long time dynamics experiment, using an independent measurement of the CDW imbalance decay. The solid line shows fit to the data using Eq.~\eqref{eq:CDW_J_fit}. The error bars are smaller than the markers. \textbf{b,} Calibration of the on-site interactions $U$ as a function of the magnetic field $B$. For each $B$, we identify the resonant staggered potential ($\Delta \approx U$) via a drop in the detected atom number (inset). The solid line shows the fit result, using Eq.~\eqref{eq:interpolation_U_B}. \textbf{c,} Spectroscopic calibration of the staggered potential $\Delta$. We calibrate the absolute energy scale of the staggered potential $\Delta$ by observing a drop in the imbalance of the initial state (inset). The error bars in the main panels (b), and (c) are the fit errors and if not visible, are smaller than the markers.}
     \label{fig:calibrations_IPR}
\end{figure*}

\subsection{Doublon splitting}
\label{sec:doublon_detection}

After the quench, we let the system evolve for a variable amount of time and then freeze the dynamics by rapidly ramping up the short lattice along the $y$-axis. To avoid parity projection of doubly-occupied sites during the fluorescence imaging, which would cause systematic errors in the evaluation of the SLE, we use the neighbouring empty chains along the $x$-axis to split doublons into singlons on two adjacent lattice sites, i.e. $\ket{2,0}\rightarrow\ket{1,1}$.

Note that this method cannot be used to detect sites with three or more atoms. We have, however, verified numerically that even in the ergodic model with $\Delta = 0$ and $U/J = 2.7(1)$, the probability of triply-occupied sites is only $\approx 1\%$, and $<0.1\%$ for higher occupancies.

As a first step of the splitting sequence after freezing the dynamics (illustrated in Fig.~\ref{fig:exp_seq}b), we ramp up the depth of the long lattice along the $x$-axis to $50\,E_\mathrm{r,l}$, simultaneously removing the DMD potential and the long lattice along the $y$-axis. At this point, the superlattice along $x$ still has a non-zero phase that we previously used for the initial-state preparation. Subsequently, we remove the short lattice along $x$ within \SI{20}{ms}. In order to ensure that the on-site interaction $U$ during the splitting step is sufficiently large, we change the magnetic field to \SI{21.7}{G}. After that we change the superlattice phase to zero in \SI{100}{ms}, corresponding to a symmetric double-well configuration. Lastly, we ramp up the short lattice along $x$ to $40\,E_\mathrm{r,s}$ in \SI{60}{ms}. The large on-site interaction energy $U$ ensures that during the splitting step, the two particles in a single well of the long lattice split between the two primary lattice sites, resulting in one atom per site.

To accurately identify the ladder systems used for doublon splitting in the raw fluorescence images, we evaluate two-point density-density correlations between neighbouring detection ladders. When the lattice partitioning is correct, no correlations are observed between neighbouring ladders, as expected. However, when the partitioning is incorrectly shifted by one lattice site, artificial correlations between neighbouring ladder systems become apparent.

\subsection{Calibration of the tunnel coupling}

To calibrate the tunnel coupling $J$, we prepare a CDW initial state and perform a quench experiment in the absence of the staggered potential $\Delta$, with the on-site interactions set to the regime of hard-core bosons. In this regime, we can use the free-fermion prediction for the imbalance decay~\cite{scherg_observing_2021, wienand_emergence_2024}
\begin{equation}\label{eq:CDW_J_fit}
    \mathcal{I}(t) = \mathcal{I}_0 \, \mathcal{J}_0\left(4(t+\delta t)\hbar/J\right).
\end{equation}
Here, $\mathcal{I}_0$ is the initial state imbalance, $\delta t$ is the time offset due to finite quench duration, and $J$ is the tunnel coupling. $\mathcal{J}_0$ denotes the 0th-order Bessel function of the first kind.

\textbf{Calibrations for the short-time dynamics (DQPT) measurements:} From the initial state data, we obtain $\mathcal{I}_0 = 0.961(1)$. Using $\delta t$ and $J$ as free fit parameters, we obtain $J/h = \SI{155(2)}{Hz}$ and $\delta t = \SI{0.07(2)}{ms}$.

\textbf{Calibrations for the long-time dynamics [$\dimH$] measurements:} From the initial state data, we obtain $\mathcal{I}_0 = 0.975(4)$. Using $\delta t$ and $J$ as free fit parameters, we obtain $J/h = \SI{122(1)}{Hz}$ and $\delta t = \SI{0.11(2)}{ms}$ (Fig.~\ref{fig:calibrations_IPR}a).

\subsection{Calibration of the on-site interaction}

The on-site interaction $U$ depends on the specific lattice parameters used for the quench experiment and on the magnetic field that controls the scattering length via the Feshbach resonance. For all quench experiments studied in this work, the long lattice along the $y$-axis was either absent, or used in the staggered configuration to introduce an on-site potential offset of the form $(-1)^i\Delta/2$, where $i$ is the site index.

For a given magnetic field and a set of lattice depths used for the quench experiment, we calibrate the on-site interaction $U$ by probing resonant doublon formation in a staggered superlattice. We start by preparing a CDW initial state and additionally use the long lattice $y$ in a staggered configuration to introduce an on-site potential offset $(-1)^i\Delta/2$, with the high-energy sites coinciding with the initially occupied sites. For sufficiently large $U$, we expect resonant formation of doublons, $\ket{\dots 101 \dots} \leftrightarrow	\ket{\dots 020 \dots}$, when $U \approx 2\Delta$. Due to parity projection, we observe the enhanced doublon creation rate at the resonance as a drop in the detected atom number (drop in Fig.~\ref{fig:calibrations_IPR}b). After performing the calibration of $U$ as a function of the staggered potential $\Delta$ for various magnetic fields, we fit the data with the following fit function, that captures the change in the $s$-wave scattering length near a Feshbach resonance (Fig.~\ref{fig:calibrations_IPR}b)
\begin{equation}\label{eq:interpolation_U_B}
    \Delta_{\text{res}}(B) = \Delta_0\left( 1- \frac{\delta_B}{B-B_0}\right).
\end{equation}
where $\Delta_0$ (prefactor), $\delta_B$ (width of the Feshbach resonance) and $B_0$ (position of the Feshbach resonance) are free fit parameters. The magnetic field corresponding to the non-interacting point, $U=0$, was further calibrated by probing the expansion of bosons in a 2D lattice~\cite{ronzheimer_expansion_2013}.

The staggered potential $\Delta$ is independently calibrated spectroscopically using a running-wave modulation. We start by preparing a CDW initial state and add a strong staggered potential $(-1)^i\Delta/2$. Here, the occupied sites coincide with the low-energy sites of the staggered superlattice. We then switch on the running-wave modulation at frequency $f$ and, after a short wait time, detect a drop in the imbalance when $\Delta \approx hf$ -- i.e., when the particles from the low-energy sites resonantly tunnel to the high-energy sites. We repeat this measurement for multiple staggered potential depths $\Delta$ (Fig.~\ref{fig:calibrations_IPR}c). For details about the running-wave implementation, see Ref.~\cite{impertro_local_2024}.

The on-site interaction strength $U/h = \SI{3.94(2)}{kHz}$ for the short-time dynamics data presented in Fig.~\ref{fig:DQPTs} has been calibrated in the same manner, this time at a fixed magnetic field of $\SI{30.2}{G}$ used for the quench experiment. This value corresponds to $U/J = 25.4(4)$, deep in the regime of hard-core bosons.

\section{Subsystem Loschmidt echo}

\subsection{Definition of the subsystem LE}

Formally, we can define the SLE using local projectors onto the initial state, averaged over all spatial subsystems
\begin{equation}
    \begin{aligned}
        \mathrm{ln}\,\mathcal{L}_N(t) &= \frac{1}{L-N+1}\sum_{i=1}^{L-N+1}\mathrm{ln} \Bigl\langle{\prod_i^{i+N-1} \hat{P}_i}\Bigr\rangle.
    \end{aligned}
    \label{eq:subsystemLE}
\end{equation}
Here, $\hat P_i$ is the local projection operator onto the initial state that acts on site $i$, and the bracket denotes the expectation value with respect to the wavefunction $|\psi(t)\rangle$.

We note that our definition generalizes to arbitrary initial product states and not only translationally invariant states investigated in the literature so far. For translationally invariant initial states, our definition simplifies to the spatial mean of $\langle{\prod_i^{i+N-1} \hat{P}_i}\rangle$, and for $N=L$, we recover the definition of the full LE. We note that it is necessary to consider subsystems with adjacent lattice sites only (see Sec.~\ref{supp_sec:local_vs_all_subsystems} for more details).

\subsection{DQPT and higher-order correlations}\label{Sec:SM_correlations}

\begin{figure*}[htb!]
\includegraphics[bb = 0 0 6.30in 2.51in,width=2\columnwidth]{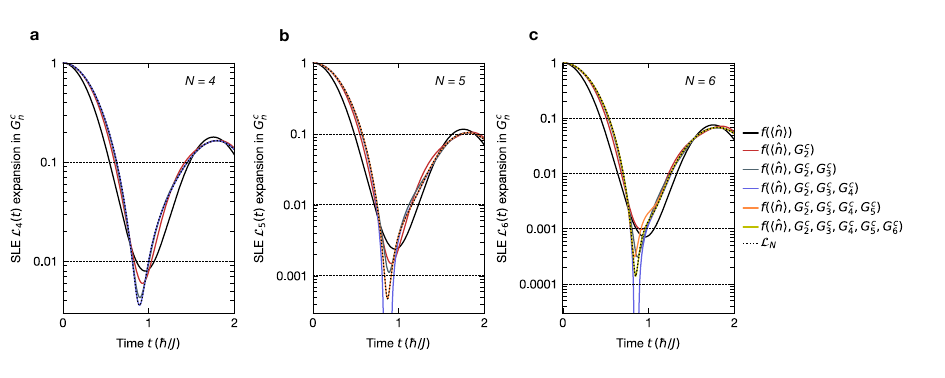}
     \caption{\textbf{Origin of the DQPT from genuine higher-order correlation functions.} \textbf{a,} Numerically evaluated contributions to the SLE expansion in Eq.~\eqref{eq:subsystemLoschmidt_L4_expansion} for subsystem size $N=4$. The simulation was performed using a TEBD algorithm for a system of size $L=26$. For the evaluation, we have considered only the central-most sites in order to avoid edge effects. \textbf{b, c,} Analogous SLE expansions for subsystem sizes $N=5$ and $6$, respectively. Black dotted line in all plots shows the directly evaluated SLE $\mathcal{L}_N$, agreeing with the result of the expansion, taking all terms into account.}
     \label{fig:suppmat:DMRG_analysis_DQPT_corrs}
\end{figure*}

In this section, we explain in more detail the expansion of the SLE in terms of higher-order correlations and provide additional numerical evidence that the DQPT arises due to non-trivial, genuine, higher-order connected correlations.

The initial state in the experiment is a CDW,  $\ket{\cdots10101010\cdots}$. We are interested in the expansion of the SLE in terms of $n$-point connected density correlation functions, such as~\cite{rispoli_quantum_2019}
\begin{equation}\label{eq:G2}
    G_2^c(i,j) = \expect{\hat n_i \hat n_j} - \expect{\hat n_i} \expect{\hat n_j},
\end{equation}
or
\begin{equation}\label{eq:G3}
    \begin{aligned}
         G_3^c(i,j,k) &= \expect{\hat n_i \hat n_j \hat n_k} -G_2^c(i,j)\expect{\hat n_k}\\ &  - G_2^c(i,k)\expect{\hat n_j} - G_2^c(j,k)\expect{\hat n_i} \\ & -  \expect{\hat n_i}\expect{\hat n_j}\expect{\hat n_k}.   
    \end{aligned}
\end{equation}

As an illustration, we will show the expansion for subsystem size $N=4$. The SLE for the CDW initial state can be expressed as
\begin{equation}\label{eq:L4_via_n}
    \mathcal{L}_4 = \expect{\hat n_0 (1-\hat n_1) \hat n_2 (1-\hat n_3)},
\end{equation}
and analogous expressions can be written for other subsystem sizes, but for odd subsystem sizes, we need to respect the unit cell, for instance, $\mathcal{L}_1 = \expect{\hat n_0 + 1 - \hat n_1}/2$. Note that because of translational invariance of the initial state, it is sufficient to look at specific sites only.

Since the initial state is translationally invariant, we can assume that $\expect{\hat n_0} = \expect{\hat n_2} = n_e$ and $\expect{\hat n_1} = \expect{\hat n_3} = n_o$. Also, the two-point density-density correlations simplify to a function of distance $d = \abs{i-j}$
\begin{equation}\label{eq:G2_into_G_d}
    G_2^c(0,1) = G_2^c(1,2) = \dots = G_2^c(d=1).
\end{equation}

The SLE for subsystem size $N=4$ can therefore be expressed as
\begin{equation}\label{eq:subsystemLoschmidt_L4_expansion}
    \begin{aligned}
        \mathcal{L}_4 &= n_e^2 (1-n_o)^2 \\ & - G_2^c(d=1)3n_e(1-n_o) + G_2^c(d=2)(n_e^2 + (1-n_o)^2) \\ & - G_2^c(d=3)n_e(1-n_o) \\ & + G_2^c(d=1)^2 + G_2^c(d=2)^2 + G_1^c(d=1)G_2^c(d=3) \\ & + n_e\left( G_3^c(0,1,3) + G_3^c(1,2,3) \right) \\ & - (1-n_o)\left( G_3^c(0,2,3) + G_3^c(0,1,2) \right) \\ & + G_4^c(0,1,2,3),
    \end{aligned}
\end{equation}
where we see contributions from the mean filling, 2-, 3- and 4-point connected higher-order correlation functions.

The experimentally evaluated contributions to the SLE, $\mathcal{L}_4$, are shown in the main text, Fig.~\ref{fig:DQPTs}d. In Fig.~\ref{fig:suppmat:DMRG_analysis_DQPT_corrs}, we further show numerical simulation results of the expansions for subsystem sizes $N = 4,5$ and $6$. The numerical results were obtained for a system of size $L=26$ using the TEBD algorithm implemented with TeNPy~\cite{tenpy2024}.

For $N=4$, we see numerically the same features as in the experimental data -- namely that the higher-order connected correlations become relevant at the dynamical critical point. For the larger subsystem sizes, $N=5, 6$, our observation about the relative importance of the higher-order correlations is even more striking. For $N=6$, for example, we indeed need to take into account also the $5$- and $6$- point correlations at the dynamical critical point to correctly capture the sharpening of the kink.

\subsection{Long-time average of the SLE}

In this section, we will discuss how the LE and the SLE in the long-time regime are related to entropies. After long evolution times, the system is expected to thermalize within the accessible Hilbert space and hence the probability to measure a particular string should be inversely proportional to the dimension of this space.

To formally prove this connection, we first analyse the full LE by rewriting the time-evolved wavefunction in terms of eigenstates of the post-quench Hamiltonian $\hat{H}$, $\ket{\psi(t)} = \sum_n c_n e^{-iE_n t}\ket{n}$, where the complex $c_n$ coefficients are given by the initial state $\ket{\psi_0} = \sum_n c_n\ket{n}$, and $E_n$ is the energy associated with eigenstate $\ket{n}$. The LE can be written as
\begin{equation}
    \begin{aligned}
        \mathcal{L}(t) = |\braket{\psi_0|\psi(t)}|^2 = \sum_{n, m} |c_n|^2 |c_m|^2 e^{-i(E_n - E_m)t/\hbar}.
    \end{aligned}
\end{equation}

Assuming the eigenstates are non-degenerate, the off-diagonal terms with  $E_n - E_m \neq 0$  average to zero when taking the time average, resulting in
\begin{equation}
    \begin{aligned}
        \overline{\mathcal{L}(t)} = \sum_n |c_n|^4 \equiv \mathrm{IPR}\equiv \exp[-S_2],
    \end{aligned}
    \label{eq:LE_vs_IPR}
\end{equation}
where ``IPR'' stands for the inverse-participation ratio and $S_2$ is the second R\'enyi entropy of the so called diagonal or time-averaged ensemble~\cite{rigol_thermalization_2008}. Note that $|c_n|^2$ represents the probability of occupying a particular energy state after the quench. The R\'enyi entropy is upper bounded by the thermodynamic entropy $S$ of the same diagonal ensemble: $S_2 \leq S$ and both can serve as somewhat different measures of the effective, accessible Hilbert space dimension: $\mathrm{dim}\,\mathcal{H}_{\mathrm{eff}}=e^S$. These measures agree when all the states within the available Hilbert space are equally populated.

Now let us analyse the time averaged SLE. To simplify the analysis, we will focus first on the homogeneous initial state, such as the CDW, and later discuss the case of inhomogeneous states, such as the domain wall. If the system is ergodic (within the accessible Hilbert space), i.e., if it satisfies eigenstate thermalization hypothesis (ETH), then the diagonal ensemble is equivalent to the thermodynamic ensemble~\cite{dalessio_quantum_2016}. Then, as we know from statistical mechanics, all subsystems of sizes $1\ll N\ll L$ can be described by the grand canonical ensemble (GCE). Note that neither the string/product observables like the one measured in the full LE, nor the R\'enyi entropy $S_2$, satisfy ensemble equivalence. For this reason, we must carefully examine the SLE in the limit $1\ll N\ll L$ without assuming that it contains the same information as the full LE. This can be done by noticing that the subsystem is described by the GCE corresponding to an inverse temperature $\beta$ and a chemical potential $\mu$, which are set by the mean energy and the number of particles in the system after the quench $\hat {\mathcal N}$. The probability to observe some state $\ket{X}$ in the subsystem is given by the expectation value of the projector to this state, $\ket{X}\bra{X}$,
\begin{equation}
    P_X=\langle X| e^{\beta (\Omega-\hat H+\mu \hat {\mathcal N})}|X\rangle,
\end{equation}
where $\Omega$ is the grand potential of the subsystem related to the partition function. In our experiment, the state $|X\rangle$ is simply the initial state $|\psi^{(i)}_0\rangle$ confined to the subsystem $\left[i, i+1, \dots, i+N-1 \right]$ with a fixed number of particles ${\mathcal N}_i=\overline{\mathcal N_i}$. Therefore, using standard thermodynamic relations, we find
\begin{equation}
    \overline {\mathcal L^{(i)}_N(t)}=e^{\beta F}\langle \psi_0^{(i)}| e^{-\beta\hat H}|\psi_0^{(i)}\rangle,
    \label{eq:susbsystem_LE_GCE}
\end{equation}
where $F$ stands for the free energy of the subsystem.

From this equation, we see that the long-time SLE is determined by the boundary partition function~\cite{heyl_dynamical_2013} which enters the time-dependent LE, where time is replaced by the temperature of the system after equilibration. The DQPTs discussed above can be viewed as a breakdown of the analytic continuation of this boundary partition function from any $\beta$ to real times beyond the dynamical critical time~\cite{heyl_dynamical_2013}. We thereby arrive at an interesting conclusion already mentioned earlier, that DQPTs can be viewed as a loss of analytic connection between short- and long-time behaviour of projectors to the initial state, similar to equilibrium phase transitions. We thus see that string observables allow us to genuinely extend the concept of finite temperature phase transitions to 1D systems, where they do not occur in standard thermodynamics.

Now, let us apply Jensen's inequality
\begin{equation}
    \langle X| e^{-\beta \hat H}|X\rangle \geq e^{-\beta \langle X|\hat H|X\rangle}=e^{-\beta E_X}
\end{equation}
to Eq.~\eqref{eq:susbsystem_LE_GCE}, leading to
\begin{equation}
 S_N^{{\rm eff},{(i)}}\equiv-\ln { \overline {\mathcal L^{(i)}_N(t)}}\leq S(\beta,\mu),
\end{equation}
where we have introduced the effective entropy of the subsystem $S_N^{{\rm eff},{(i)}}$, which is extracted from the experiments and the standard thermodynamic entropy $S(\beta,\mu)$. We thus see that the SLE provides a lower bound to the thermodynamic entropy. Interestingly, this is not in contradiction to the result for the full LE, as the R\'enyi entropy is also lower-bounded by $S$. The estimate of the difference between $S_N^{{\rm eff},{(i)}}$ and the thermodynamic entropy comes from applying the lowest order cumulant expansion to the boundary partition function $\langle X| e^{-\beta \hat H}|X\rangle\approx e^{-\beta E_X-(\beta^2/2) \delta E_X^2}$, where $\delta E_X^2=\langle X| \hat H^2|X\rangle-\langle X| \hat H|X\rangle^2$ is the energy variance. From this we find that
\begin{equation}
    S_N^{{\rm eff},{(i)}} \equiv -\ln { \overline {\mathcal L^{(i)}_N(t)}}\approx S(\beta,\mu)-{1\over 2}{\beta^2 \delta E_i^2},
\end{equation}
We see that the correction is small either in the high-temperature limit or when the quench results in a small energy variance.

Finally let us comment on what happens for an inhomogeneous initial state. The analysis applies equally, except that in Eq.~\eqref{eq:susbsystem_LE_GCE}, we will pick up an additional term $\beta \mu ({\mathcal N_i}-\bar {\mathcal N}_i)$, where ${\mathcal N_i}$ is the actual number of particles in the subsystem $i$ while $\bar {\mathcal N}_i$ is the mean number of particles in this subsystem. Similarly, for the effective entropy inequality, we will pick up extra terms
\begin{equation}
    S_N^{{\rm eff},{(i)}}\leq S(\beta,\mu)+\beta (\bar E_i-E_i)-\beta \mu ({\mathcal N_i}-\bar {\mathcal N}_i).
\end{equation}
Interestingly, summing over $i$, all these extra contributions cancel and we again obtain the full entropy bound for the SLE as in the homogeneous case.

We note that the connection between the thermodynamic entropy and $\mathrm{ln}\,\mathcal{L}_N$ motivates the choice to average spatially $\mathrm{ln}\,\mathcal{L}_N^{(i)}$, and not $\mathcal{L}_N^{(i)}$.

\subsection{Dimension of the accessible Hilbert space}

Let us now explain the connection between the thermodynamic entropy, $S$, and the dimension of the accessible Hilbert space. As we have understood in the previous section and in the main text, we can view the subsystem as being connected to a grand-canonical bath -- i.e., based on measuring subsystems of finite size, we directly extract properties in the thermodynamic limit, $L\to\infty$. Once the subsystem size $N$ is larger than the correlation length $\xi$ in our measurement basis (i.e., real-space density measurement), $\text{d}S_N^{{\rm eff}}/\text{d}N$ becomes approximately constant. This means, that we can use the extensive entropy to extrapolate to the thermodynamic limit, $L \to \infty$, as
\begin{equation}
    \mathrm{dim}\,\mathcal{H}_{\mathrm{eff}} = \mathrm{exp} \left( L\diff{S_N^{{\rm eff}}}{N} \right),
\end{equation}
directly leading to Eq.~\eqref{eq:slope_vs_dimH}.

\begin{figure}[t]
    \centering
    \includegraphics[bb = 0 0 3.50in 4.12in, width=1\columnwidth]{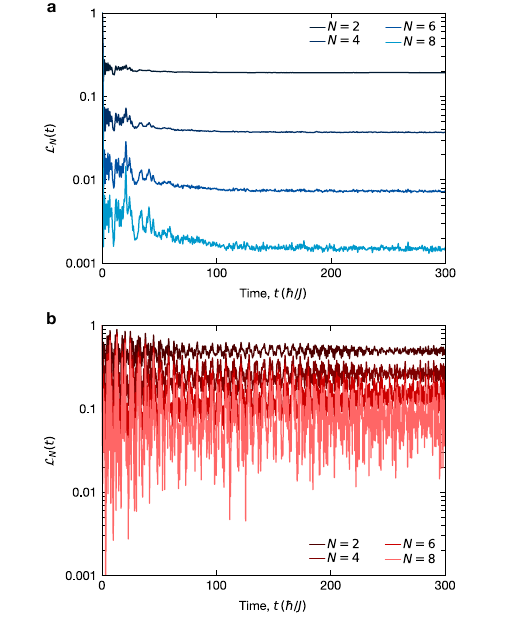}
    \caption{\textbf{Relaxation time of the SLE.} \textbf{a,} SLE following a quench in the ergodic regime ($U/J=2.7$, $\Delta=0$), starting with a CDW initial state. \textbf{b,} SLE following a quench in the fragmented regime ($\Delta/J=16.7$, $\Delta = U$), starting with a single-domain-wall initial state. Numerical results were obtained using the Krylov subspace method in a system of $L = 16$ sites.}
    \label{fig:suppmat_relaxation_time}
\end{figure}

\begin{figure}[t]
    \centering
    \includegraphics[]{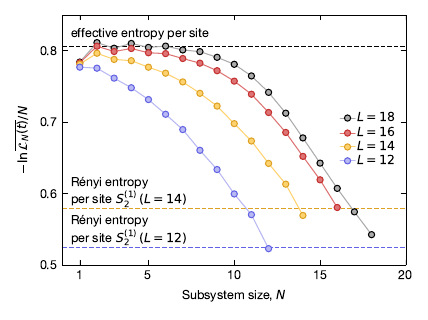}
    \caption{\textbf{SLE, the effective entropy, and the R\'enyi entropy.} The effective entropy density $S^{\mathrm{eff}}_N/N$, where ${S_N^{\mathrm{eff}} \equiv -\mathrm{ln}\,\overline{\mathcal{L}_N}}$, saturates to a value greater than the R\'enyi entropy per site $S_2/L$, where $L$ is the system size. This is consistent with the inequality $S_2 \leq S$, where $S=S_{vN}$ is the thermodynamic entropy per site, and our theory that the SLE for $1\ll N\ll L$ saturates to the thermodynamic entropy of the subsystem. We note that the observed saturation value of the SLE is quite close to the analytical \textit{von Neumann entropy} per site $\approx 0.90$, obtained by assuming at most two bosons per lattice site. As $N\rightarrow L$, the SLE value smoothly approaches the R\'enyi entropy. Numerical results were obtained using the Lanczos method of unitary time evolution at a fixed Krylov space dimension of 20. The R\'enyi entropy was calculated using exact diagonalization.}
    \label{fig:suppmat_vn_vs_Renyi_entropy}
\end{figure}

\subsection{Relaxation times and saturation of the SLE}

In order to extract the accessible Hilbert space dimension, we need to make sure that the SLE has relaxed to a steady-state value. In Fig.~\ref{fig:suppmat_relaxation_time}, we plot the numerically calculated SLE for subsystem sizes $N = 2,4,6$ and $8$ in the ergodic regime ($U/J = 2.7$, CDW initial state) and in the fragmented regime ($\Delta/J = 16.7$, $\Delta = U$, domain wall initial state). We observe that with increasing subsystem size $N$, the SLE takes longer and longer to reach relaxation. For the time-window we have looked at in the experiment, and the relatively small subsystem sizes $N\leq 5$ we have probed, we can see that the SLE has indeed relaxed.

Next, in Fig.~\ref{fig:suppmat_vn_vs_Renyi_entropy}, we numerically verify that the saturation value $S^{\mathrm{eff}}_N$ of the SLE is indeed given by the thermodynamic entropy $S_N$ of the subsystem, unlike the time-averaged rate function, that saturates to the R\'enyi entropy $S_2$. For intermediate subsystem sizes, $S^{\mathrm{eff}}_N/N$ saturates closer to the von Neumann entropy per site $S_{vN}^{(1)} \approx \ln{2.46}\approx 0.90$, which is analytically obtained in the thermodynamic limit by assuming at most two atoms per lattice site [see the discussion following Eq.~\eqref{eq:sm_vn_entropy}]. As $ N\rightarrow L $, and SLE approaches the full LE, we measure the R\'enyi entropy $S_2$ as expected.

\subsection{Robustness of the SLE}\label{sec:SIII.C}

\begin{figure}[t]
    \centering
    \includegraphics[]{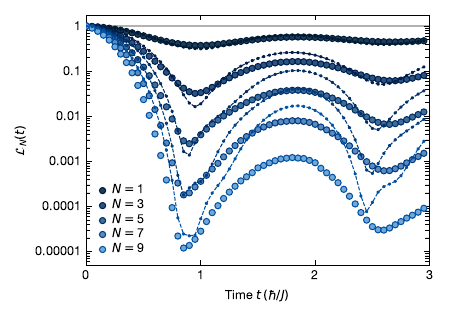}
    \caption{\textbf{Robustness of the SLE against noise.} Rate-function of the SLE as a function of time shows the deviation of noisy dynamics (large markers) from the clean dynamics (dashed lines). The SLE becomes more susceptible to noise for larger $N$. The noisy dynamics is simulated by decreasing the average filling with $2\%$ probability at each time step (i.e., an exaggerated scenario of atom loss in the experiment). Simulation was done for a CDW initial state, system size $L=14$, and  $U/J\approx 24$. Each time point is averaged over 200 realizations.}
    \label{fig:noisy_dqpt}
\end{figure}

In this section, we compare the robustness of finite SLE strings against noise to show that the SLE is much more stable than the full LE. To simulate a noisy setup, we consider a finite probability of particles randomly escaping the lattice (thus breaking the conservation of particle number), which can, in reality, reflect atom loss in the experiment. It is immediately clear that measuring the full LE would yield zero even if one particle is lost. However, the SLE is stable against such events. In Fig.~\ref{fig:noisy_dqpt}, we show the numerically calculated short-time dynamics of the SLE, assuming an exaggerated atom loss scenario. We observe that the SLE for smaller $N$ is more robust against atom loss, and the deviation of the noisy data from perfect dynamics increases for larger $N$. This suggests that even if one is able to measure the full LE, for large system sizes, it is extremely susceptible to noise, and it is indeed more favourable to experimentally probe the SLE.

Furthermore, from Fig.~\ref{fig:noisy_dqpt}, it is clear that atom loss during the dynamics washes away any sharp non-analytic feature in the rate function, and this effect is, of course, more pronounced at later times as more particles are lost. In the numerical simulation, we cut off the particle loss once the filling fraction falls below $0.2$ so that we do not hit the trivial vacuum state.

\subsection{SLE with all versus local subsystems}\label{supp_sec:local_vs_all_subsystems}

\begin{figure}[t]
    \centering
    \includegraphics[bb = 0 0 3.50in 4.57in, width=1\columnwidth]{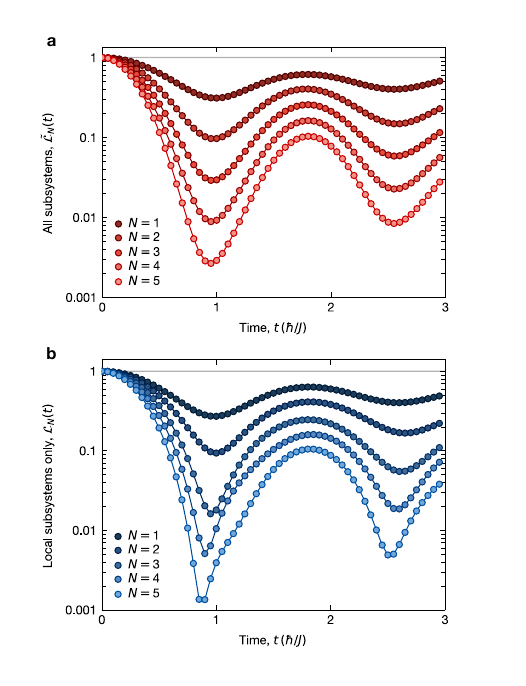}
    \caption{\textbf{SLE evaluated using all subsystems versus local subsystems only.} \textbf{a,} When considering all connected and disconnected subsystems, the rate-function is always analytic for all $N$. \textbf{b,} When consider only locally connected subsystems, we observe a gradual emergent sharpening of the kink in the rate function with increasing $N$. Both simulations were performed for the CDW initial state, system size $L=20$, $U/J = 24$ and $\Delta=0$.}
    \label{fig:loc_all}
\end{figure}

In order for the SLE to capture the properties of the LE, particularly the non-analytic features in its rate function, the SLE needs to be calculated using locally connected subsystems only. In Fig.~\ref{fig:loc_all}, we compare the numerically evaluated SLE for local subsystems only (Fig.~\ref{fig:loc_all}b) and the SLE evaluated using \textit{all} including non-local subsystems (Fig.~\ref{fig:loc_all}a), i.e., connected and disconnected ones. Considering all subsystems, there is no emergent sharpening of the non-analyticity and no DQPT. Considering local connected subsystems only, the DQPT in the SLE emerges.

This observation is entirely consistent with our observation in the main text, that the genuine higher-order connected correlations are instrumental in the appearance of the DQPT. Although these are still included when considering all subsystems, this approach puts more weight on few-body long-range connected correlations compared to the highest-point correlations for a given subsystem size. This, in turn, smoothens out the rate-function, and it fails to capture true non-analyticities in the limit $N\rightarrow L$.

\section{Data analysis}

\subsection{Atom loss, post-selection and ROI}

As the first step of data analysis, we reconstruct single-site occupations from the raw fluorescence images~\cite{impertro_unsupervised_2023}. We then select a large region of interest (ROI) within our box potential and filter out a small fraction of fluorescence images that contain few or no atoms due to errors in the sequence execution. Each fluorescence image contains $20-40$ independent copies of the system. As a second step, we post-select chains by filling in a specified range around half-filling. This step is necessary to minimize the impact of imperfect initial state and (potentially chain-dependent) particle loss during the time evolution.

\subsubsection{Short-time dynamics}

During short-time dynamics, we do not detect any atom loss in the ROI of size $32\times 32$ within a box potential of $40\times 40$ lattice sites. During post-selection, we filter out approximately $20\%$ of the chains with filling lower than $0.42$ and higher than $0.55$.

\subsubsection{Long-time dynamics}

As we have described above, for the long-time dynamics, we prepare the initial state only in every other chain, leaving the remaining half of the chains empty for subsequent doublon detection to avoid parity projection during the fluorescence imaging. We again post-select on chains that have a filling less than $0.42$ and above $0.55$. This procedure filters out up to roughly $40\%$ of the chains. For the CDW, we have used a ROI of $28\times 28$ in a box containing $30\times 30$ sites. For the domain wall, we have used an ROI of $52\times 36$ sites in a box containing $60\times 40$ sites.

In Table~\ref{table:atomloss}, we summarize the measured mean filling for both the initial states and the late-time data between approximately $(230-280)\hbar/J$ that are analysed in Fig.~\ref{fig:IPR}b of the main text. The initial state filling for the data shown in Fig.~\ref{fig:IPR}c was $0.52(3)$. The mean-filling after the dynamical evolution was for all $\Delta/J$ values above $0.42(2)$ (before post-selection on individual chain filling). The CDW filling has been evaluated in a ROI of $28\times 28$ in a box containing $30\times 30$ sites. The domain-wall filling has been evaluated in a ROI of $52\times 36$ sites in a box containing $60\times 40$ sites.

\begin{table}
\centering
\begin{tabular}{| c | c | c |}
\hline
 mean filling & CDW & domain wall \\ 
 \hline
 initial state & $0.49(3)$ & $0.51(3)$ \\  
 ergodic regime & $0.49(2)$ & $0.43(1)$ \\
 fragmented regime & $0.50(2)$ & $0.43(2)$ \\
 \hline
\end{tabular}
\caption{\textbf{Atom loss data.} Atom loss data for the long-time regime measurement shown in Fig.~\ref{fig:IPR}b. The numbers are evaluated after post-selection on failed shots, but before post-selecting chains based on their filling. We have used the large ROI for the evaluation: $28\times 28$ sites for the CDW initial state and $52\times 36$ for the domain wall initial state.}\label{table:atomloss}
\end{table}

For the data analysis, the same ROI was used, except in the case of the domain wall in the ergodic regime. In this special case, due to to a long relaxation time of the domain wall mass imbalance, we use a smaller ROI $52\times 20$ centred on the domain wall. The residual imbalance we measure within the ROI is defined as $\mathcal{I} = \left(n_{\mathrm{o}} - n_{\mathrm{e}}\right) / \left(n_{\mathrm{o}} + n_{\mathrm{e}}\right) $, where $n_{\mathrm{o}}$ is the mean filling on the initially occupied sites and $n_{\mathrm{e}}$ is the mean filling on the initially empty sites sites. The impact of the ROI size on the extracted $\left(\mathrm{dim}\,\mathcal{H}\right)^{1/L}$, together with the residual imbalance, is detailed in Table~\ref{table:roi_domainwall}. Going to ROI smaller than $20$ lattice sites becomes challenging due to limited subsystem statistics.

\begin{table}
\centering
\begin{tabular}{| c | c | c |}
\hline
 ROI size & $\left(\mathrm{dim}\,\mathcal{H}\right)^{1/L}$ & $\mathcal{I}$ \\ 
 \hline
 $20$ & $2.29(1)$ & $\approx 6\%$  \\  
 $30$ & $2.24(1)$ & $\approx 10\%$ \\
 $40$ & $2.19(1)$ & $\approx 15\%$ \\
 \hline
\end{tabular}
\caption{\textbf{Effect of the ROI size on the evaluation in Fig.~\ref{fig:IPR}b.} Effect of the ROI size on the evaluation for the domain wall initial state in the ergodic regime, plotted in Fig.~\ref{fig:IPR}b.}\label{table:roi_domainwall}
\end{table}

\subsection{Subsystem Loschmidt echo evaluation}\label{suppmat:SLE_exp_eval}

We note that the SLE is evaluated by assuming a perfect initial state, ignoring imperfections in the initial state preparation and possible atom loss during the dynamics.

In the experiment, we obtain spatially resolved return probabilities $\mathcal{L}_N^{(i)}$, where $i$ is the spatial index of a subsystem. Since in the long-time regime the local SLE corresponds to the local effective entropy, we average spatially over $\mathrm{ln}\,\mathcal{L}_N^{(i)}$. This is in contrast to averaging $\mathcal{L}_N^{(i)}$ directly - this would, for instance, lead to the rate function no longer being an intensive quantity. However, for the special case of translationally invariant initial states, such as the CDW, $\mathcal{L}_N^{(i)}$ is approximately constant across the lattice, and the two averages coincide.

We note that experimentally, given a limited number of snapshots, spatially averaging $\mathrm{ln}\,\mathcal{L}_N^{(i)}$ is more challenging, since we have to ensure that we have enough samples to resolve non-zero values of $\mathcal{L}_N^{(i)}$ for every spatial subsystem, the reason being that the logarithm of zero is not well defined.

\textbf{Short-time dynamics.} The SLE for the short-time dynamics (Fig.~\ref{fig:DQPTs}) was calculated by spatially averaging the SLE  $\mathcal{L}_N^{(i)}$. The error bar on the SLE was calculated from the standard error of proportion, since the SLE is nothing else than a probability to detect a rare event (an initial-state bitstring). Given $N_{\mathrm{meas}}$ measurements (in our case, the total number of subsystems from all the snapshots for a given evolution time) and the probability of an event $p$, the standard error of proportion is given by $\sqrt{\frac{p(1-p)}{N_{\mathrm{meas}}}}$. When we detect $N_{\mathrm{events}} = pN_{\mathrm{meas}}$ successful events, the standard error on this number is approximately $\sqrt{N_\mathrm{events}}$. This is, however, nothing else than a standard deviation for a process governed by a Poisson distribution.

\textbf{Long-time dynamics.} The SLE for the long-time measurement (Fig.~\ref{fig:IPR}) was calculated by spatially averaging $\mathrm{ln}\,\mathcal{L}_N^{(i)}$, since the domain wall initial state is not translationally invariant and we are evaluating entropies.

The error bars were estimated using a bootstrapping procedure, where we sample chains from the dataset, each chain being a single snapshots of $\ket{\psi(t)}$. We repeat the sampling 2000 - 5000 times.

We note that for the bootstrapping procedure we draw a small number of samples that, for the largest subsystem size $N=5$, return zero SLE $\mathcal{L}_N^{(i)}$ for at least a single spatial subsystem, and thus zero SLE for the whole snapshot, since spatially averaging $\mathrm{ln}\,\mathcal{L}_N^{(i)}$ is equivalent to taking a geometric mean over the chain. We discard these samples but have verified that the error bars are not strongly affected by this procedure.

\subsection{Extraction of the DQPT sharpness}

\begin{figure*}
    \centering
    \includegraphics[bb = 0 0 6.30in 2.15in,width=2\columnwidth]{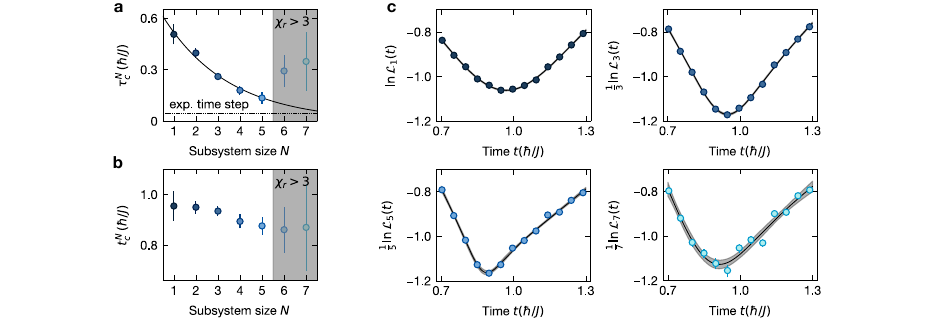}
    
     \caption{\textbf{Extraction of the characteristic time and the location of the DQPT.} \textbf{a,} Fitted characteristic time $\tau_c^N$ for subsystem size $N$. The solid line shows an exponential fit to the data for $N = 1 - 5$. The shaded region indicates that for the two largest subsystem sizes, $N=6,7$, $\chi_r > 3$. \textbf{b,} Fitted location of the DQPT, $t_c^N$. \textbf{c,} Fit to the experimental data for $N = 1, 3, 5$ and $7$. The shaded region shows the $\pm1\sigma$ confidence interval of the fit.}
     \label{fig:char_time_fit}
\end{figure*}

In order to quantitatively extract the sharpness of the SLE DQPT signal for a subsystem of size $N$, we fit an empirical fit function to the experimentally measured rate function $\mathrm{ln}\,\mathcal{L}_N/N$ (Fig.~\ref{fig:DQPTs}b). The empirical function describes a piece-wise linear fit ($\alpha_N(t-t_c^N) + c$ for $t<t_c^N$, $\beta_N(t-t_c^N) + c$ for $t>t_c^N$) with a smooth cross-over region described by a tanh function with characteristic time $\tau_c^N$ and offset $c$,
\begin{equation}
    \begin{aligned}
        \mathcal{L}_N^{\rm fit}(t) &= \alpha_N (t-t_c^N) \frac{1-\mathrm{tanh}\left( \frac{2(t-t_c^N)}{\tau_c^N} \right)}{2} \\ & + \beta_N (t-t_c^N)\frac{1+\mathrm{tanh}\left( \frac{2(t-t_c^N)}{\tau_c^N} \right)}{2} + c.
    \end{aligned}
    \label{eq:peak_fit}
\end{equation}

For an ideal DQPT captured by the LE in the thermodynamic limit, $L\to\infty$, we expect a perfect piece-wise linear dependence close to the peak, i.e. $\tau_c^{L\to\infty} = 0$.

The fit results are shown in Fig.~\ref{fig:char_time_fit}. In Fig.~\ref{fig:char_time_fit}a, we show the fitted characteristic time $\tau_c^N$, together with an exponential fit to the data. In Fig.~\ref{fig:char_time_fit}b, we show the extracted location of the DQPT $t_c^N$. In Fig.~\ref{fig:char_time_fit}c, we show example fits to the rate function for $N = 1,3, 5$ and $7$.

In the main text, we refrain from showing the characteristic time for the two largest subsystem sizes, $N=6, 7$, since the data point spacing is not fine enough for us to reliably capture the increasing sharpness, and because the signal-to-noise ratio for the two largest subsystems gets worse. Indeed, based on the exponential fit in Fig.~\ref{fig:char_time_fit}a, we would expect the characteristic time $\tau_c^{7}$ to be comparable with the spacing between the data points. This fact also reflects in reduced chi-square $\chi_r$ of the fit, which becomes larger than $3$ for $N=6, 7$.

\subsection{DQPT numerics with imperfect initial state}

\begin{figure}[t]
    \centering
    \includegraphics[]{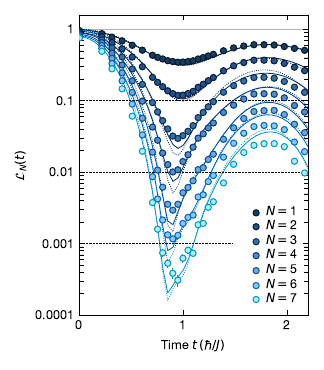}
    \caption{\textbf{Effect of imperfect initial state.} Comparison between experimental and numerical data for the SLE, $\mathcal{L}_N$. The solid lines show imperfect initial state numerics, and the dotted lines show numerics for a perfect initial state.}
    \label{fig:DQPT_exp_vs_num}
\end{figure}

We have also performed numerical simulations to study the effect of initial state imperfections on the SLE. In the experiment, filled sites have an occupation probability of approximately $93\%$ and empty sites have an occupation probability of $2\%$. In Fig.~\ref{fig:DQPT_exp_vs_num}, we compare the numerics for the imperfect initial state and the experimental data shown in Fig.~\ref{fig:DQPTs}. Including the initial state imperfections indeed leads to a better agreement with the experimental data. Each numerical data point was averaged over $200$ independent dynamical evolutions for a system size of $L=20$ using the Lanczos method of unitary time evolution at a fixed Krylov space dimension of 20.

\section{Effective Hilbert space dimension and fragmentation}

In this section, we derive the effective Hamiltonian for the 1D Bose-Hubbard model (BHM) with strong staggered potential. The Hamiltonian realized in the experiment can be expressed as
\begin{equation}
    \begin{aligned}
        \hat H &= - J \sum_{i} \left(  \hat a_{i}^{\dagger} \hat a_{i+1}^{\phantom{\ast}} + \mathrm{h.c.} \right)\\ & + \underbrace{\frac{U}{2} \sum_{i} \hat n_{i} (\hat n_{i}-1)}_{\hat{H}_U} + \underbrace{\frac{\Delta}{2} \sum_{i} (-1)^i \hat n_{i}}_{\hat{H}_\Delta}.
    \end{aligned}
    \label{eq:bh}
\end{equation}
Here, $\hat{a}^{\dagger}$ ($\hat{a}$) represents the bosonic creation (annihilation) operator, $\hat{n}_i$ is the number operator, $J$ denotes the tunnel coupling, $\hat{H}_U$ captures the on-site interactions, and $\hat{H}_\Delta$ the staggered potential.

\textbf{Ergodic model.} Before discussing the Hilbert space fragmentation (HSF), let us take a look at the dimension of the accessible Hilbert space in the ergodic model. For the non-interacting BHM at half-filling ($\Delta/J = 0$ and $U/J = 0$), the Hilbert space dimension can be calculated using simple combinatorics (stars and bars method) as
\begin{equation}
\mathrm{dim}\,\mathcal{H}_{\mathrm{BHM}} = \binom{\frac{3}{2}L - 1}{L-1}, 
\label{d_hf}
\end{equation}
with $L$ denoting the system size. Using Stirling's approximation, for $L \to \infty$ one obtains
\begin{equation}
\mathrm{dim}\,\mathcal{H}_{\mathrm{BHM}} \approx 2.60^L
\end{equation}

Alternatively, we can calculate the \textit{von Neumann entropy} in the \textit{grand-canonical ensemble}. Using textbook arguments, the probability $p_k$ that a single lattice site is occupied by $k$ non-interacting bosons is $p_k = z^k/\Xi$, where $z = e^{-\beta\mu}$ and $\Xi = \sum_{k=0}^{\infty} z^k = 1/(1-z)$ is the grand canonical partition function. The chemical potential is fixed by the filling: $\overline{n} = \sum_{k=0}^{\infty} k p_k = z/(1-z) = 1/2$. Thus, $z=1/3$, and we find that the von Neumann entropy per site is
\begin{equation}\label{eq:sm_vn_entropy}
S_{vN}^{(1)} = - \sum_{k=0}^\infty p_k \mathrm{ln}\,p_k = \beta\mu\overline{n} - \mathrm{ln}\,\Xi \approx \mathrm{ln}\,2.60
\end{equation}

Similarly, allowing at most two bosons per lattice site, we get $\Xi = 1 + z + z^2$. At half-filling, $z = (-1 + \sqrt{13})/6$, and we find the von Neumann entropy per site $S_{vN}^{(1)} \approx \mathrm{ln}\,2.46$.

\textbf{Fragmented model.} Let us now discuss Hilbert space fragmentation (HSF). We consider a model with strong staggered potential in the resonant regime where $\Delta \approx U$ and $U \gg J$. In this case, $\langle \hat{H}_{U} + \hat{H}_{\Delta}\rangle \approx \mathrm{const.}$, and any hopping process has to conserve $\langle \hat{H}_{U} + \hat{H}_{\Delta}\rangle$. Thus, the allowed processes are $|11\rangle_{i,i+1} \leftrightarrow |20\rangle_{i,i+1}$ for odd spatial index $i$, and $|11\rangle_{i,i+1} \leftrightarrow |20\rangle_{i,i+1}$ for even spatial index $i$. In the limit of large $\Delta$, the effective first-order Hamiltonian can therefore be written as
\begin{equation}
    \begin{aligned}
        \hat{H}_{\text{frag}} &= -J \left( \sum_{\text{odd}\,i} \left( |11\rangle_{i,i+1}\langle 20|_{i,i+1} + \text{h.c.} \right) \right. \\
        &\quad \left. + \sum_{\text{even}\,i} \left( |11\rangle_{i,i+1}\langle 02|_{i,i+1} + \text{h.c.} \right) \right).
    \end{aligned}
    \label{eq:h_frag}
\end{equation}

We consider three initial states, illustrated below for $L=12$:

\begin{enumerate}
    \item the CDW initial state $|010101010101\rangle$,
    \item the single-domain-wall (sDW) initial state $|111111000000\rangle$ and
    \item the double-domain-wall (dDW) initial state $|000111111000\rangle$.
\end{enumerate}

\begin{figure}[t]
    \centering
    \includegraphics[bb = 0 0 3.50in 1.96in, width=1\columnwidth]{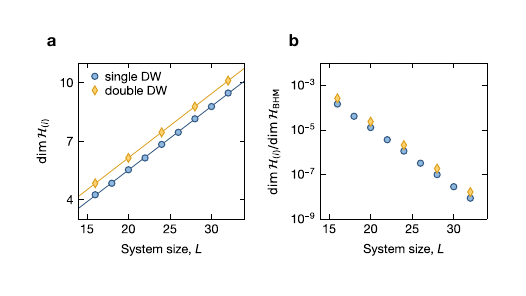}\\
  \caption{\textbf{a,} Dimensions of fragments corresponding to single-domain-wall initial state $|\psi_{0}^{\rm sDW}\rangle$ and double-domain-wall initial state $|\psi_{0}^{\rm dDW}\rangle$ as a function of system size $L$. Solid and dashed lines show linear fits to the data, both yielding a slope of $\approx 0.33$. \textbf{b,} The ratio $\mathrm{dim}\,\mathcal{H}_{\mathrm{sDW/dDW}}/\mathrm{dim}\,\mathcal{H}_{\mathrm{BHM}}$ approaches zero with increasing system size $L$.}\label{s1}
\end{figure}

Under the effective Hamiltonian defined in Eq.~\eqref{eq:h_frag}, the CDW initial state is a completely frozen state, and the dimension of the corresponding fragment is equal to 1. For the single- and double-domain-wall initial states, we can directly calculate the dimensions of the fragments by counting the number of allowed configurations. The results of this calculation are shown in Fig.~\ref{s1}a. We note that the double-domain-wall state corresponds to the largest fragment.

Using simple combinators, we can calculate the dimension of the single-domain-wall fragment under the effective Hamiltonian as
\begin{equation}
\mathrm{dim}\,\mathcal{H}_{\mathrm{sDW}} = \binom{L/2}{L/4}.
\label{d_analytic_frag}
\end{equation}
Applying Stirling's formula, we obtain
\begin{equation}
    \mathrm{dim}\,\mathcal{H}_{\mathrm{sDW}} \approx \sqrt
{2}^L
\end{equation}
As shown in Fig.~\ref{s1}b, it is clear that for the double-domain-wall initial state,
\begin{equation}
    \frac{\mathrm{dim}\,\mathcal{H}_{\mathrm{sDW/dDW}}}{\mathrm{dim}\,\mathcal{H}_{\mathrm{BHM}}} \rightarrow 0,
\end{equation}
indicating strong HSF under the effective Hamiltonian.

\begin{figure}[t!]
    \centering
    \includegraphics[bb = 0 0 3.50in 3.35in, width=1\columnwidth]{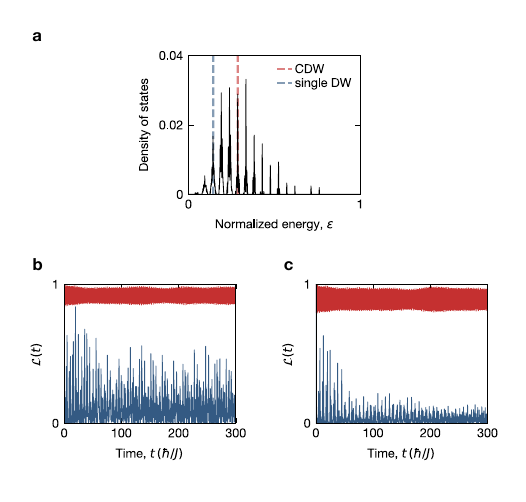}\\
  \caption{\textbf{a,} Density of states (DoS) as a function of the normalized energy $\epsilon$ for staggered BHM in the fragmented regime ($U=\Delta$, $\Delta/J = 16.7$). Simulation done for system size $L=12$ using exact diagonalization. The dashed vertical lines highlight the normalized energies $\epsilon$ of the CDW and single-domain-wall initial states. \textbf{b,} Time evolution of the LE for the CDW and single-domain-wall initial states, simulated using the Krylov subspace method. Same parameters as in (a), system size  $L=12$. \textbf{c,} Same as (b), but with larger system size $L=16$.}\label{s2}
\end{figure}

To further explore our fragmented model, we use exact diagonalization (ED) to numerically calculate the density of states (DoS) for a system of size $L=12$ at half-filling. In Fig.~\ref{s2}a, we show the DoS as a function of normalized energy $\epsilon = (E-E_{\text{min}})/(E_{\text{max}} - E_{\text{min}})$, with $E_{\text{min(max)}}$ being the minimum (maximum) energy of the Hamiltonian $\hat{H}$. The DoS shows multiple pronounced peaks corresponding to different fragments. In the same Figure, we also highlight the normalized energy $\epsilon$ for the CDW and the single-domain-wall states. Thus, we ensure that the dynamics corresponding to the two initial states is of non-equilibrium nature in the sense that these initial states have energies far beyond the ground state, and lie in the regions with large non-zero DoS.

Next, in Fig.~\ref{s2}b, we plot the dynamics of the LE for the staggered BHM in the fragmented regime ($U=\Delta$ and $\Delta /J=16.7$). The system size was chosen to be $L=12$. For the CDW initial state, the LE exhibits an almost frozen dynamics, oscillating around a mean value close to 1. In contrast, for the single-domain-wall initial state, the LE oscillates around a much smaller mean value. For larger system size, $L=16$, we observe the same behaviour of the dynamics (Fig.~\ref{s2}c).

We note that even though the DoS of the CDW state is larger than that of the single DW initial state, the LE for the single DW state has a lower value, corresponding to a larger accessible Hilbert space dimension. This suggests that this phenomenon is different from the many-body mobility edge~\cite{guo_observation_2021}.

\begin{figure}[t]
    \centering
    \includegraphics[]{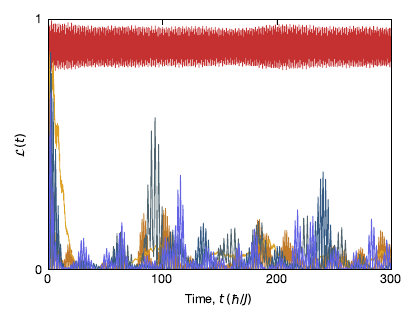}
    \caption{Dynamics of the LE in the fragmented regime ($U=\Delta$, $\Delta/J = 16.7$), simulated using the Krylov subspace method. The red line shows the dynamics for the approximately frozen CDW initial state. The remaining lines show dynamics of the LE for different initial states with the same particle number and energy. System size is $L=16$.}\label{s_add1}
\end{figure}

We further study the dynamics of LE with five additional initial states with the same particle number and energy as the CDW initial state. The results are shown in Fig.~\ref{s_add1}. Despite having the same particle number and energy, the additional states show very different dynamical behaviour.

The time evolution of the LE in Fig.~\ref{s2}b, c and Fig.~\ref{s_add1} is numerically simulated using the Krylov subspace methods.

\section{Long-time regime numerics}

In this section we present numerical simulations of the long-time dynamics of the SLE, and compare our results to the experimental data. We start by presenting numerical results for a perfect initial state and then extend them to capture experimental imperfections, such as imperfect initial state preparation and atom loss during the dynamics.

\subsection{Numerical results for perfect initial state}

\begin{figure}[t!]
    \centering
    \includegraphics[bb = 0 0 3.50in 1.53in, width=1\columnwidth]{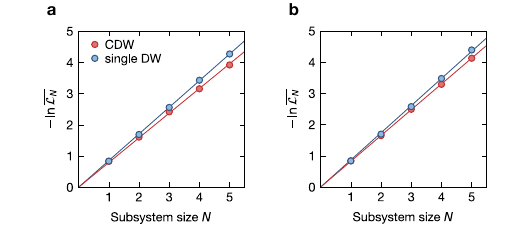}\\
  \caption{\textbf{SLE numerics in the ergodic regime.} \textbf{a,} Numerical results for the time-averaged free energy density $-\ln \overline{\mathcal{L}_{N}}$ as a function of the subsystem size $N$ in the ergodic regime ($U/J=2.7$, $\Delta/J=0$) for the CDW and single-domain-wall initial states. We fit $\dimH = 2.21(1)$ for the CDW and $2.35(1)$ for the sDW, respectively. Simulation was done using the Krylov subspace method. Here, the system size is $L=12$. The solid lines show linear fits to the numerical data (forced to go through zero). \textbf{b,} Numerical results for $L=16$. We fit $\dimH = 2.28(1)$ for the CDW and $2.39(1)$ for the sDW, respectively.}\label{fig:suppmat:SLEergodicNumerics}
\end{figure}

In Fig.~\ref{fig:suppmat:SLEergodicNumerics} we show the numerical results for the time-averaged rate function $-\ln \overline{\mathcal{L}_{N}}$ as a function of subsystem size $N$, both for the CDW and single-domain-wall initial states. The simulation was done using the Krylov subspace method for two different system sizes, $L=12, 16$, and we took spatial average of $\mathrm{ln}\,\mathcal{L}_N^{(i)}$. The time-window for the averaging was chosen to be $(100 - 300)\hbar/J$. Using a linear fit function (forced to go through zero), we extract the dimension of the accessible Hilbert space $\left(\mathrm{dim}\,\mathcal{H}_{\mathrm{eff}}\right)^{1/L}$. For the single DW initial state we obtain $\dimH = 2.35(1)$ and $2.39(1)$ for $L=12$ and $16$ respectively. For the CDW initial state we obtain $\dimH = 2.21(1)$ and $2.28(1)$ for $L=12$ and $16$ respectively.

\begin{figure}[t!]
    \centering
    \includegraphics[bb = 0 0 3.50in 2.90in, width=1\columnwidth]{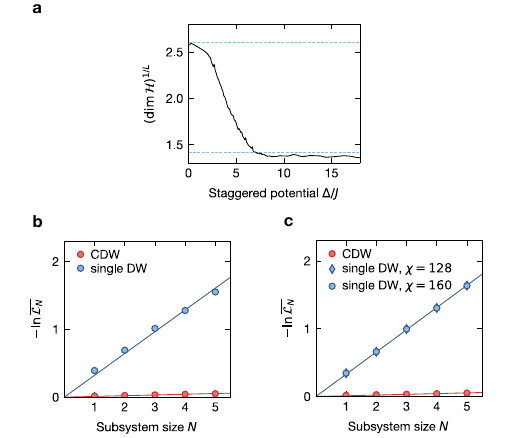}\\
  \caption{\textbf{SLE numerics in the fragmented regime.} \textbf{a,} Numerically extrapolated dimension of the accessible Hilbert space $(\text{dim} \mathcal{H})^{1/L}$ as a function of varying strength of the kinetic constraints $\Delta/J$, keeping $\Delta = U$. The system size for the simulation was $L=16$ and the simulation was performed using the Krylov subspace method. The dashed green line shows the analytic prediction for non-interacting bosons, $\approx 2.60$, and the blue dashed line the prediction based on the effective Hamiltonian, $\approx 1.41$. \textbf{b,} Numerical results (Krylov subspace method) for the time-averaged free energy density $-\ln \overline{\mathcal{L}_{N}}$ as a function of the subsystem size $N$ in the fragmented regime ($\Delta/J=16.7$, $U=\Delta$), simulated for CDW and single-domain-wall initial states and system size $L=16$. Solid lines show the results of linear fits used to extract the Hilbert space dimension. We extract $\dimH = 1.01$ for the CDW and $1.38(1)$ for the sDW, respectively. \textbf{c} Same as (b), but for significantly larger system size $L=60$, simulated using the TDVP algorithm. Note that the linear fit for the single DW initial state 
  works significantly better for the larger system size. We extract $\dimH = 1.01$ for the CDW and $1.39(1)$ for the sDW, respectively.}\label{fig:suppmat:SLEfragmentedNumerics}
\end{figure}

Next, we investigate the effect of varying strength of the staggered potential $\Delta/J$ on the dimension of accessible Hilbert space, keeping $U=\Delta$. We repeat the dynamical evolution simulation for different values $\Delta/J$ between 0 and 18, and for each $\Delta/J$ extract the dimension of accessible Hilbert space $\left(\mathrm{dim}\,\mathcal{H}_{\mathrm{eff}}\right)^{1/L}$. The results for system size $L=16$ are shown in Fig.~\ref{fig:suppmat:SLEfragmentedNumerics}a. For $\Delta/J \simeq 0$, the extracted dimension is that for non-interacting bosons at half-filling and at infinite temperature temperature ($\approx 2.60^L$, indicated by the green dashed line). With increasing depth of the staggered potential we observe a cross-over regime and for $\Delta/J \gtrsim 10$ we observe the emergence of a plateau, where the dimension of the accessible Hilbert space no longer changes. There, we find $\left(\mathrm{dim}\,\mathcal{H}_{\mathrm{eff}}\right)^{1/L}$ consistent with the combinatorics prediction based on the effective Hamiltonian in the fragmented regime ($\approx \sqrt{2}^L$, indicated by the blue dashed line). The numerical simulation is also qualitatively consistent with experimental data presented in Fig.~\ref{fig:IPR}c of the main text.

In Fig.~\ref{fig:suppmat:SLEfragmentedNumerics}b, we show the time-averaged rate function $-\ln \overline{\mathcal{L}_{N}}$ in the fragmented regime for both CDW and single DW initial states. The simulation was done for $\Delta/J = 16.7$ and system size $L=16$. Using a linear fit function, we extract $\dimH = 1.38(1)$ for the single DW initial state and $\dimH \approx 1.01$ for the CDW initial state.

The numerical data for $L=16$ show a visible bending from the linear fit. In order to verify whether this bending arises due to finite system size effects, we repeat the simulation for significantly larger system size, $L=60$, using a matrix-product-state-based algorithm -- time-dependent variational principle (TDVP)~\cite{fishman_itensor_2022}. We consider a cutoff of the dimension of bosonic operators with the maximum 4, and two different values of the bond dimension, $\chi=128$ and $160$, to test the convergence.

The results of the simulation for $L=60$ are shown in Fig.~\ref{fig:suppmat:SLEfragmentedNumerics}c. Indeed, for the large system size we observe an excellent agreement with the linear fit, confirming that the bending observed in Fig.~\ref{fig:suppmat:SLEfragmentedNumerics}b was due to finite system size effects. For the large system, we extract $\dimH = 1.39(1)$ for the single DW initial state and $\dimH \approx 1.01$ for the CDW initial state.

\subsection{Effect of experimental imperfections}

\begin{figure}[t!]
    \centering
    \includegraphics[bb = 0 0 3.50in 2.94in, width=1\columnwidth]{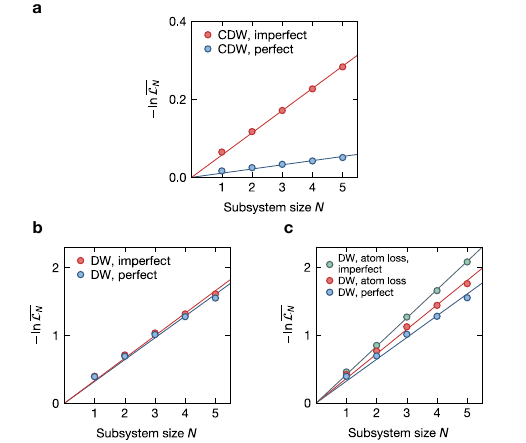}\\
  \caption{\textbf{Effect of imperfections on the SLE.} \textbf{a,} Numerical results (Krylov subspace method) for the time-averaged free energy density $-\ln \overline{\mathcal{L}_{N}}$ as a function of the subsystem size $N$ in the fragmented regime ($\Delta/J=16.7$, $U=\Delta$), simulated for CDW initial state and system size $L=16$, both with and without initial state imperfections. For the perfect initial state we fit $\dimH \approx 1.01$ and for the imperfect one $\approx 1.06$. \textbf{b,} Same as (a) but for the single-domain-wall initial state. For the perfect initial state we fit $\dimH = 1.38(1)$ and for the imperfect one $1.39(1)$. \textbf{c,} Numerical simulation for the single-domain-wall initial state, including the effect of atom loss, and combining the effects of atom loss and imperfect initial state. Using the linear fit, we extract $\dimH = 1.38(1)$ for the perfect initial state, $1.43(1)$ taking the atom loss into account, and $1.52(1)$ taking into account both atom loss and initial state imperfections. For comparison, we also plot the data without taking imperfections into account.}\label{fig:suppmat:SLEimperfectNumerics}
\end{figure}

In this section, we aim to explain the residual difference between the experimentally observed $\dimH$ in the fragmented regime, and the numerical predictions for a perfect initial state. For all numerical results in this section, we use system size $L=16$ and $\Delta/J = 16.7$, keeping $U = \Delta$. 

We start by considering the effect of initial state imperfections. We assume that there is a 5\% (1\%) error rate to prepare the state $|1\rangle$ ($|0\rangle$) on a given lattice site. Numerically, we average the SLE over 100 imperfect initial states, performing the time-dynamics simulation for each of them separately. We plot the result for the CDW initial state in Fig.~\ref{fig:suppmat:SLEimperfectNumerics}a. Taking the initial state imperfection into account, we extract $\dimH \approx 1.06$, larger than $\dimH \approx 1.01$ we extract in the perfect case, and closer to the experimentally measured value.

We also check the influence of the imperfect initial state on the single-domain-wall initial state (see Fig.~\ref{fig:suppmat:SLEimperfectNumerics}b). Interestingly, the imperfections lead only to a small difference compared to the perfect initial state case. Thus, we also include the effect of atom loss (see Table~\ref{table:atomloss}), which is stronger for the domain wall initial state than it is for the CDW initial state. We model the atom loss in the following way: we take an atom occupation snapshot $(n_{1},n_{2},...,n_{L})$, and randomly change $n_{i}$ to $n_{i}-1$ ($i\in \{1,2,...,L\}$) if $n_{i}>0$ with a probability of $10\%$. We then calculate the SLE based on these modified snapshots.

In Fig.~\ref{fig:suppmat:SLEimperfectNumerics}c, we show the SLE, $-\ln \overline{\mathcal{L}_{N}}$, for the perfect case, taking into account the atom loss, and taking into account the atom loss together with initial state imperfections. The extracted slope increases in each step, and for the simulation taking into account both the atom loss and the imperfect initial state, we obtain $\dimH = 1.52(1)$, very close to the experimentally measured value.

\begin{center}
\textbf{SUPPLEMENTARY REFERENCES}
\end{center}
\vspace{0.5em}

\putbib[manuscript]
\end{bibunit}